\documentclass[review]{article}
\usepackage[multiple]{footmisc}
\usepackage{epsfig}
\usepackage{verbatim}
\usepackage{amssymb}
\usepackage{amsmath}
\usepackage{graphicx}
\usepackage{subfig}
\usepackage{lineno}
\usepackage[top=2cm, bottom=2cm, left=2.5cm, right=2.5cm]{geometry}

\begin{document}

\title{Characterization of 3 mm Glass Electrodes and Development of RPC Detectors for $INO-ICAL$ Experiment}

\author{Daljeet Kaur, Ashok Kumar, Ankit Gaur, Purnendu Kumar,\\ Md. Hasbuddin, Swati Mishra, Praveen Kumar, Md. Naimuddin\footnote{Corresponding author: nayeem@cern.ch} \\
\\
Department of Physics and Astrophysics, University of Delhi,\\ Delhi 110007, India. }



\maketitle

\begin{abstract}
India-based Neutrino Observatory (INO) is a multi-institutional facility, planned to be built up in South India. The INO facility will host a 51 kton magnetized Iron CALorimeter (ICAL) detector to study atmospheric muon neutrinos. Iron plates have been chosen as the target material whereas Resistive Plate Chambers (RPCs) have been chosen as the active detector element for the ICAL experiment. Due to the large number of RPCs needed ($\sim$ 28,000 of $2~m \times 2~m$ in size) for ICAL experiment and for the long lifetime of the experiment, it is necessary to perform a detailed $R\&D$ such that each and every parameter of the detector performance can be optimized to improve the physics output. In this paper, we report on the detailed material and electrical properties studies for various types of glass electrodes available locally. We also report on the performance studies carried out on the RPCs made with these electrodes as well as the effect of gas composition and environmental temperature on the detector performance. We also lay emphasis on the usage of materials for RPC electrodes and the suitable enviormental conditions applicable for operating the RPC detector for optimal physics output at INO-ICAL experiment.
\end{abstract}
\vspace{1cm}
~~~~~~~~ {Keywords: INO, ICAL, RPC, GLASS }

 \newpage

\section{Introduction}\label{sec:intro}
The India-based Neutrino Observatory (INO)~\cite{ino} is a multi-institutional collaboration aimed at building a world-class underground laboratory with an all around rock cover of approximately 1 km for non-accelerator based high energy and nuclear physics experiments. It will be a multi-experiment facility and one of the experiments will be Iron CALorimeter (ICAL). The ICAL detector will have modular structure having three modules with each module weighing about 17 ktons and 16 m $\times$ 16 m $\times$ 14.5 m in dimension. The entire detector will be magnetized using a magnetic field of around 1.5 Tesla to detect the charge and momenta of particles passing through it. The primary goal of INO-ICAL is to study the atmospheric muon neutrinos. Neutrinos are fundamental particles belonging to the lepton family in the standard model of particle physics. They come in three flavours, one associated with electrons and the others with their heavier partners, the muons and the taus. According to standard model of particle physics neutrinos are massless. However, recent experiments \cite{sno}-\cite{circa} indicate that neutrinos have finite but small masses that are yet to be precisely measured. Their flavours also changes as they travel, which is known as flavour oscillation. Determination of neutrino masses and mixing parameters is one of the unresolved problems in physics today. The INO-ICAL experiment with its massive magnetized detector will be able to shed light on this issue by measuring the atmospheric oscillation parameters and neutrino mass hierarchy. In order to make these measurements, INO-ICAL needs a robust detector system through which the muons generated due to the interaction of neutrinos with the target material in ICAL can be detected easily. The Resistive Plate Chamber (RPC) detectors are one such detector which can be used as an active detector element in INO-ICAL experiment.

\section{Resistive Plate Chamber Detector}\label{sec:rpc}
The high energy and astroparticle physics experiments generally require detectors which are stable, having long lifetime and cover large detection area. The RPC detector \cite{Santonico1981,Santonico1988}, is one such gaseous detector which was pioneered during 1980s. The RPCs detectors are made up from two high resistive parallel plates with uniform graphite coating on opposite sides, generating a uniform electric field when a bias voltage is applied on the two plates. These are rugged, low cost and highly efficient detectors which have their usage in various 
high energy, nuclear and astroparticle physics experiments as well as applications in imaging and other fields. 
The main features of RPC detectors are their large pulse heights, excellent timing resolution better than few nanoseconds, high detection efficiency for minimum ionising particles and low cost per unit area. Their excellent time resolution, possibility of high granularity readout and good detection efficiency make them ideal for the development of fast, efficient and robust triggering system for muons over large areas in hybrid detectors~\cite{Bencze}. These detectors are also used for time of flight measurements~\cite{Santonico1988,BlancoTOF} and for tracking in multilayer configurations in BELLE~\cite{BELLE}, BaBar~\cite{BaBar}, L3~\cite{L3}, OPERA~\cite{OPERA}, and the LHC experiments (ALICE, ATLAS and CMS)~\cite{ALICE} - \cite{CMS}. RPC detectors has the ability to provide two dimensional readout at a time. These are also very simple to construct and are almost free from damaging discharges.

In INO-ICAL experiment, RPCs are proposed to be used as active detector element. Their high gain, simple design and affordable cost makes it favourite to construct about 28,000 detectors of dimension, 2 m $\times$ 2 m, that INO-ICAL need.  The experiment is planned to operate at least for 20 years and may also be upgraded to 100 kton in later phases. Because of the huge number of detectors required for INO-ICAL and keeping in mind the long life span of the experiment, it is pertinent to perform a vigorous $R\&D$ to carefully optimize the various detector design and operational parameters like the electrode material, gas composition, operational conditions, etc. to exploit fully all the advantages of the RPC detectors. Keeping in view the huge quantity of glasses required and the complexity 
involved in its long distance transportation, we decided to pursue our $R\& D$ on the 
glasses available locally. The earlier $R\&D$ performed for INO-ICAL glass RPCs were for 2 mm thick glasses only and was also limited in its scope~\cite{Pramana,Proceeding}. Due to handling issues and shear stress problems involved in the large size detectors, INO-ICAL collaboration has decided to use 3 mm thick glasses for constructing the RPCs, therefore, all the results reported in this paper are for 3 mm thick glasses.

\subsection{Electrodes choice for RPC}\label{sec:electrodes}
In RPC detectors, material properties of electrodes and environmental factors play a significant role in detector working. The electrodes are usually made up of float glass or bakelite and the choice of it depends on the expected event rate. Electrode material plays an important role in the functioning of the RPC detectors, so its selection must be done carefully. Typically, the electrodes should have high resistivity to control 
counting rate and preventing spread of discharge throughout the entire gas volume~\cite{Santonico1981, Czyrkowski, Cardarelli1993}. It must also have uniform surface  to avoid localization of excess charge and to prevent 
alternating leakage path during the avalanche process. 

Since INO-ICAL experiment is going to construct huge number of RPCs which in turn requires a large amount of the electrode's material, therefore, one factor that also becomes very important in the selection of electrodes is the ease of availability of it at a reasonable cost. 
We selected three types of glasses \textit{viz}., Saint Gobain, Asahi and Modi which are easily and readily available in the local market at a reasonable price for our $R\& D$.  


We performed the bulk resistivity and surface resistivity measurements of the electrodes to determine their electrical properties. The uneven surface texture of electrodes surface properties gives a profound effect on the detector performance because of the increase in dark current, counting rate and thus influencing the efficiency of the detector. For the purpose of determining the surface properties we performed Scanning Electron 
Microscopy (SEM) and Atomic Force Microscopy (AFM) analysis. 

\subsubsection{Bulk and Surface Resistivity}
\label{sec:bulkRes}
\begin{figure}
\centering
\includegraphics[height=8cm,width=10cm]{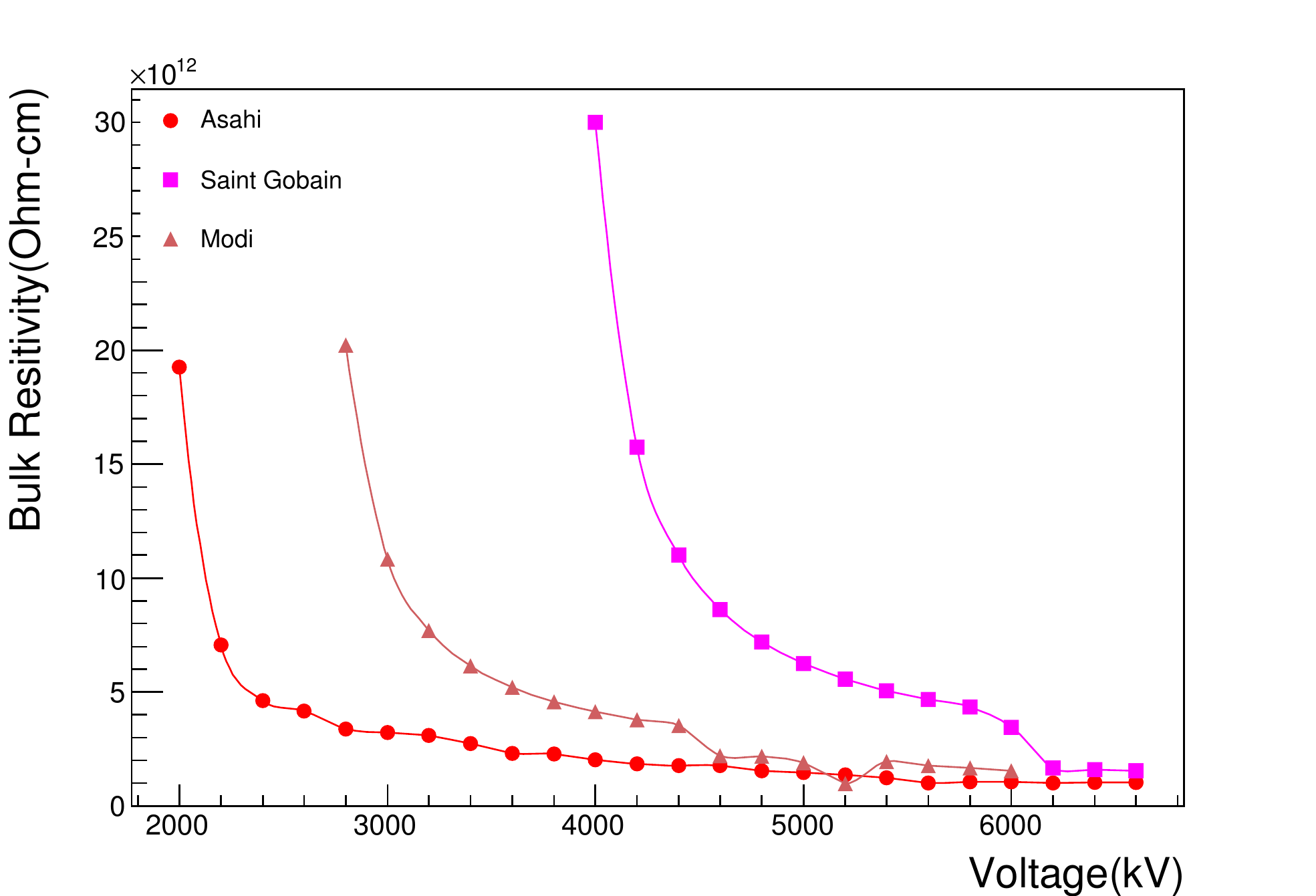}
\caption{ Bulk Resistivity of all the three types of glasses in $\Omega$-cm as a function of the applied voltage.}
\label{fig:BulkR}
\end{figure}

The bulk resistivity of the electrode material of the RPC was determined through
the measurement of current through the glass for a given bias voltage. The bulk resistivity has been measured using the glass samples cut into 3~cm $\times$ 3~cm of size. Two cables were soldered on two different copper plates which are connected to the CAEN N471A high voltage power supply.  
The copper plates were placed on the adjacent flat surfaces of the glass sample and a bias voltage was applied across it. The current measured by the inbuilt ammeter inside CAEN N471A was noted down as a function of the applied voltage. Fig.~\ref{fig:BulkR} shows the bulk resistivity of 
the three types of glasses as a function of the applied bias voltage. The Saint Gobain glass was found to have the 
maximum bulk resistivity among all the three samples. In the operating region of around 6 kV, the Saint Gobain bulk resistivity is of the order $5\times 10^{12}$ $\Omega$-cm on average. Asahi and Modi were found to be having similar bulk resistivity and is approximately $1.5 \times 10^{12}$ $\Omega$-cm in the operating region around 5 kV. The bulk resistivity measurements were repeated many times and the results varied within less than 5\% from its mean value. 

\begin{figure}
\subfloat[Asahi]{\includegraphics[width=0.33\linewidth]{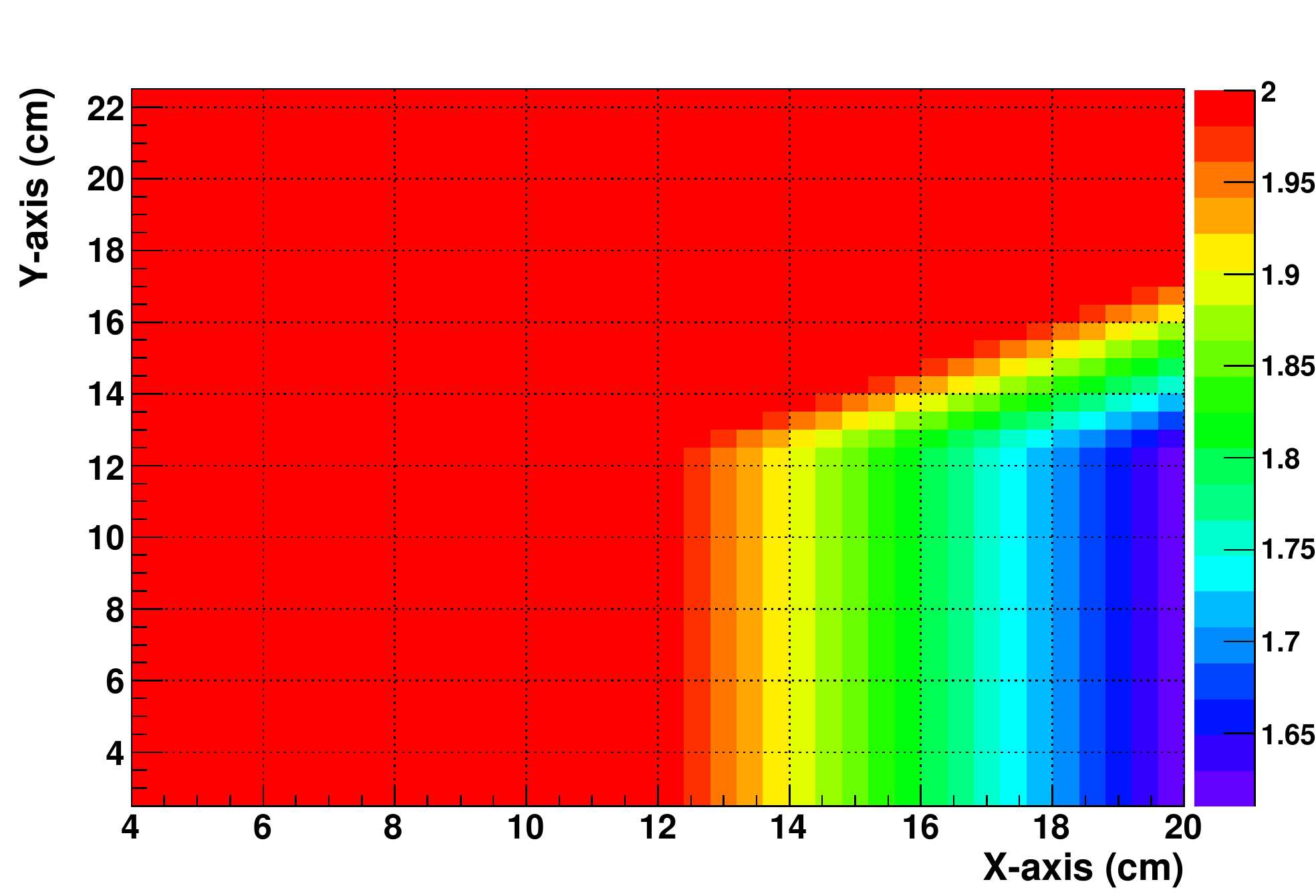}}
\subfloat[Saint Gobain]{\includegraphics[width=0.33\linewidth]{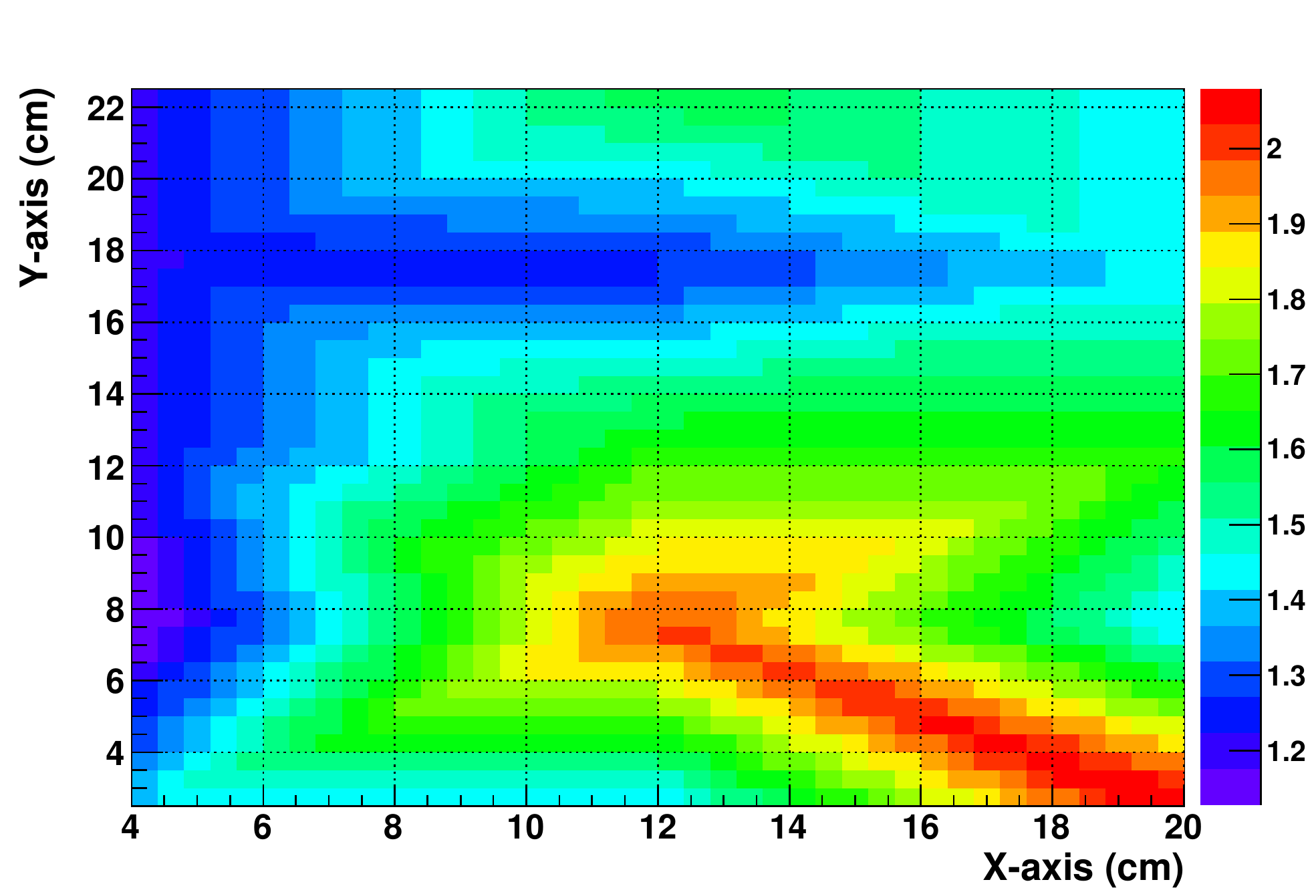}}
\subfloat[Modi]{\includegraphics[width=0.33\linewidth]{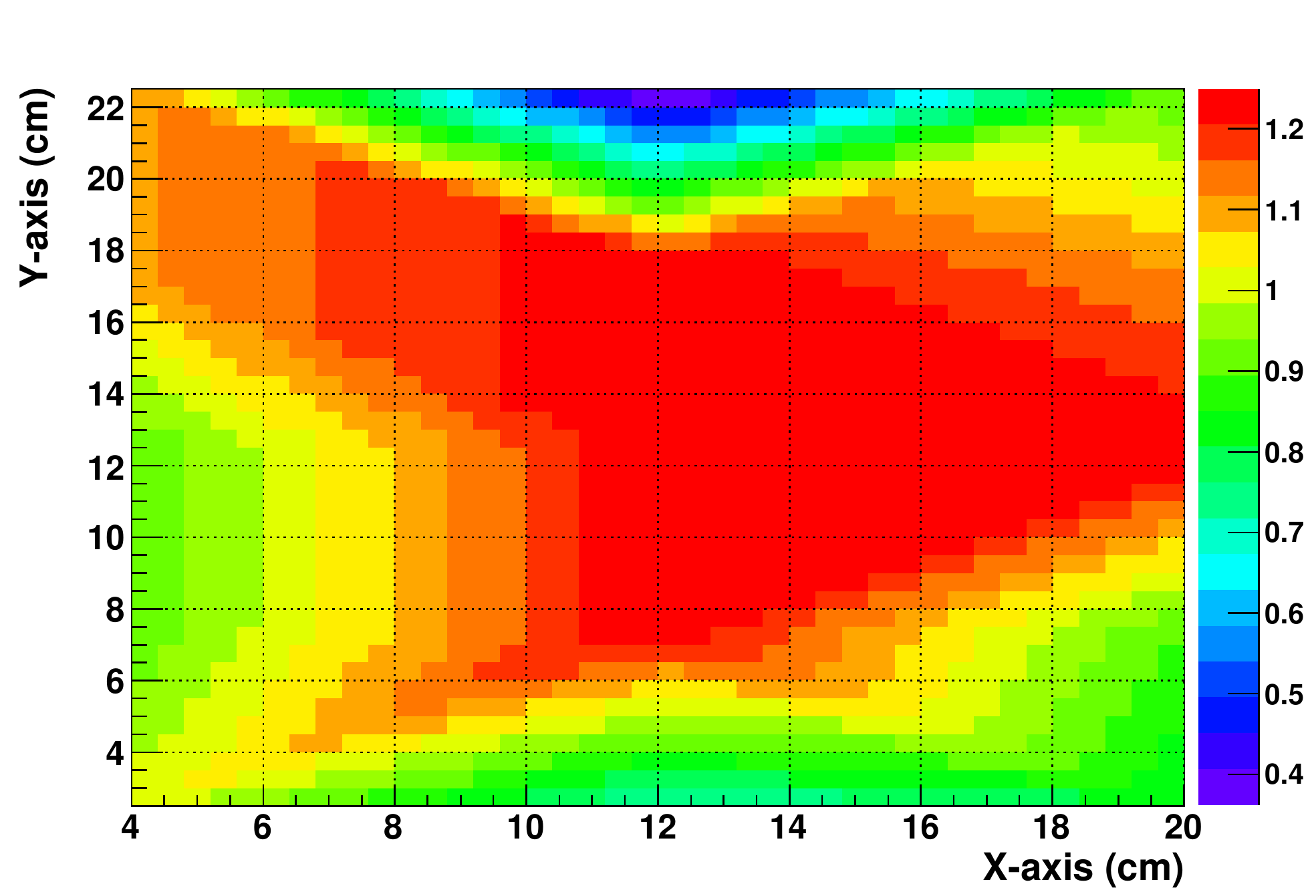}}
\caption{Surface Resistivity for all the three glasses. The X and Y axes are the length and breadth in centimeters of the glass plates whose surface resistivity was measured. The intensity of the colours indicates the value of the surface resisitivity in $10^{11}\Omega/\Box$ and the variation in colours represents the variation in the surface resistivity at different points across the glass plate.}
\label{fig:SurfaceR}
\end{figure}

The circuit for surface resistivity measurement is made with two brass bars with soft padded conducting edges at the bottom, which were placed on the test sample. The length of the brass bars and their separation were kept at 5 cm. A DC bias voltage was applied on the jig and the leakage current across the terminals of the jig through the sample was measured. A total of 10 measurement points were taken on each plate to obtain a fine binning. Also, the measurements were repeated many times for the same point and the results were found to be reproducible within 8\% uncertainty from the mean value. Fig.~\ref{fig:SurfaceR} shows the average surface resistivity measurements of the three glasses. It can be seen from this figure that the surface resistivity of the Asahi glass is approximately constant over most of surface area, whereas there are significant variations in the surface resistivity distribution of the Saint Gobain glass. The Modi glass has even wider surface resistivity variations. From the scale of surface resistivity variations of the three glasses, it can be concluded that the Asahi is having the smoothest surface of all followed by Saint Gobain and Modi respectively~\cite{Saikat_SR}. The measurments of bulk and surface resistivity were performed under normal pressure, relative humidity maintained between 45\% to 50\% and temperature between $20^{\circ}$C to $21^{\circ}$C.



\subsubsection{Surface Properties}\label{sec:surfaceProp}
\begin{figure}
\subfloat[Asahi]{\includegraphics[width=0.33\linewidth]{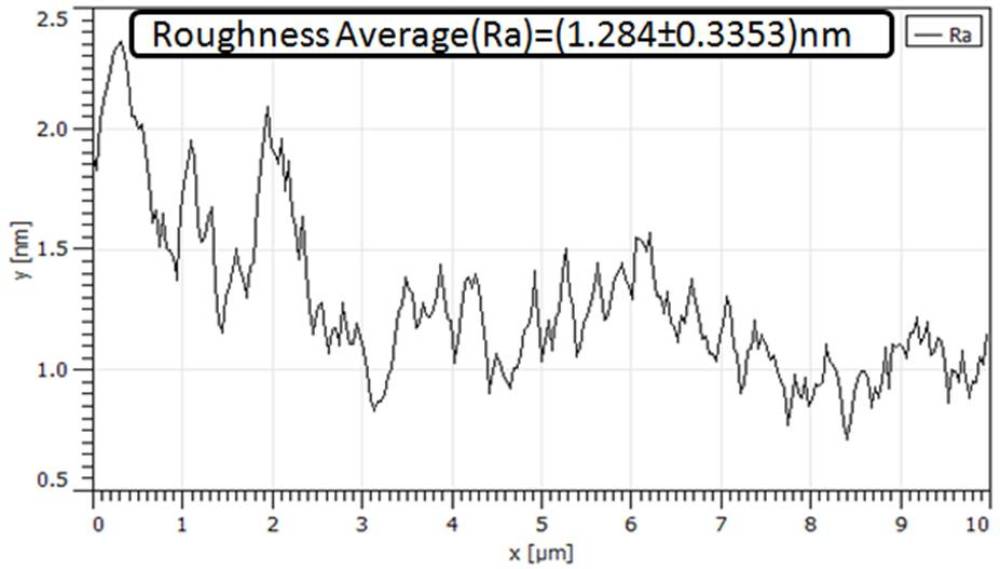}}
\subfloat[Saint Gobain]{\includegraphics[width=0.33\linewidth]{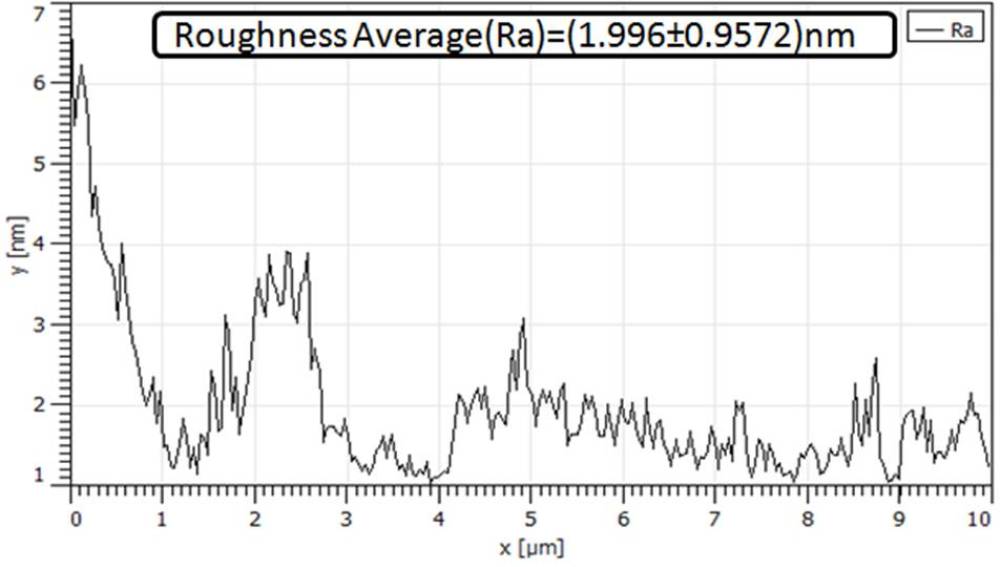}}
\subfloat[Modi]{\includegraphics[width=0.33\linewidth]{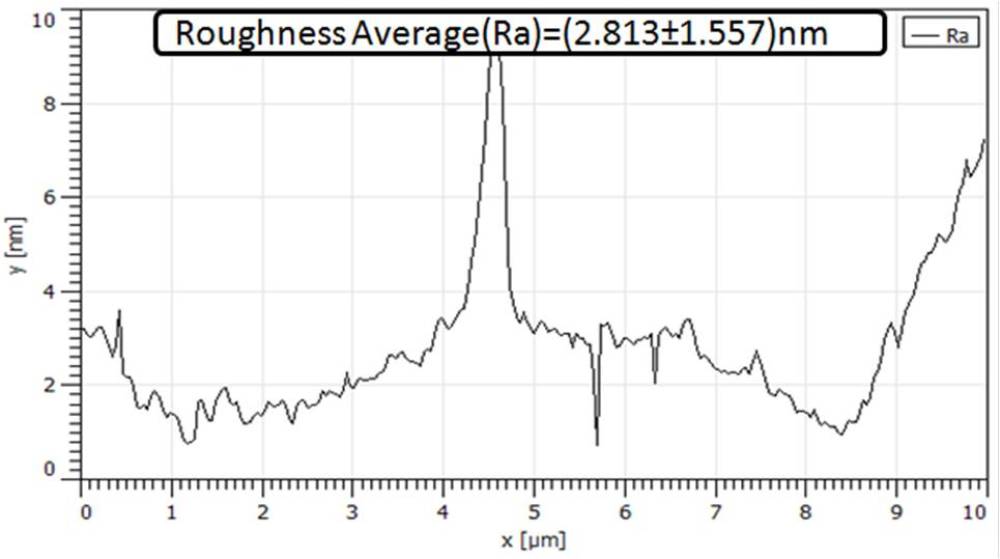}}
\caption{Atomic Force Microscopy scans for the three glass samples. The X-axis shows the scanned length of the glass sample in $\mu$m, and Y-axis shows the variation in the surface roughness (or smoothness) in $nm$}
\label{fig:AFM}
\end{figure}

\begin{figure}
\centering
\includegraphics[height=6cm,width=15cm]{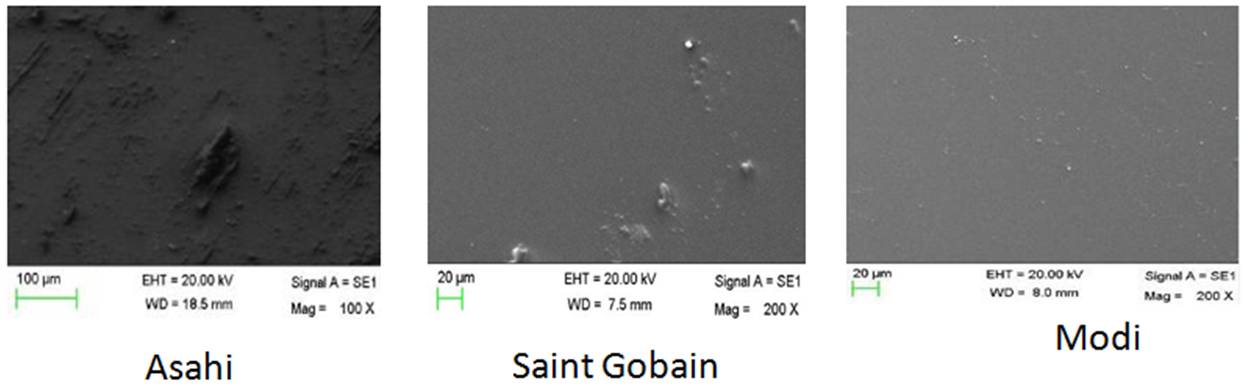}
\caption{ Electron Microscopy scans for the three glass samples.}
\label{fig:SEM}
\end{figure}

The rough inner surface of the electrodes may cause wide variation in the electric 
field inside the RPC. Due to rough surface the sharp uneven edges in the surface morphological structure causes the electric field inside the RPC to vary significantly. Rough surface is very sensitive to the field emission which is a source of high dark current and high counting rates. Therefore, it is very important to perform a panoramic study of surface roughness for the efficient operation of the RPC. We determined the roughness (or smoothness) of the 
electrodes surface though the AFM scanning. The average roughness is defined as:
\begin{equation}
R_{av} = \sqrt {\frac {\sum\limits_{i} (t_{av}-t_i)^2}{N}}
\end{equation}
where, $t_{av}$ is the average height, $t_i$ is the individual measured height as shown in Fig.~\ref{fig:AFM}, and N is the number of data points scanned on the surface. The average roughness of $1.28\pm 0.33$ nm, $2.00\pm 0.96$ nm and $2.81\pm 1.56$ nm was found for Asahi, Saint Gobain and Modi respectively. The errors on the roughness are standard errors on the mean value. The SEM images for the three glasses are shown in Fig.~\ref{fig:SEM}. The SEM images were taken from ZEISS MA15 instrument at 20 kV using common electron beam. The AFM and SEM measurements alone are not conclusive enough to declare any sample as smoother than other with high confidence, however, after taking into account the surface resistivity results together with the AFM and SEM measurements, we can conclude that, at least, qualitatively Asahi glass appears to have the smoothest surface of all. Since these results were not conclusive enough quantitavely to reject any of the three glass samples at this stage, we proceeded to fabricate the RPCs using all the three samples to perform the characterization studies.

To confirm that the glasses we have selected are amorphous in nature, we performed an X-ray diffractometer scan with REGAKU D-2000 instrument with acceleration voltage of 40 kV, current of 100 mA, and $Cu-K\alpha$ as an X-ray. 
The XRD study confirms molten glassy type matrix for all the samples. We also performed the study for the determination of the composition percentage of elements present in electrode samples. Carbon, Sodium, Magnesium, Aluminium, Silicon, Calcium, Oxygen and Tin have been found in different quantity in the three glasses. The detailed composition is given in Table ~\ref{tab:EDS}. It can be seen from the table that Saint Gobain glass has no Carbon component whereas Asahi and Modi have 4.89\% and 6.68\% Carbon component respectively. Since Carbon is a conducting element so absence of it from the Saint Gobain glass could be one of the reasons for its high bulk resistivity campared to Asahi and Modi.

\begin{table} [ht] 
\centering 
\begin{tabular}{|c|c|c|c|c|}
\hline
Serial No. & Element Name  & \multicolumn{3}{|c|}{Percentage of Elements}   \\ 
\hline 
         &          & Asahi & Saint Gobain & Modi   \\

\hline 
 1 &     C &4.89	 & -	& 6.68\\
 2 &     Na & 6.89 &	8.22 &	6.34\\
  3&     Mg & 1.79	& 1.85	& 1.55\\
  4&     Al & 0.32	& 0.27	& 0.31\\
  5&     Si & 21.7	& 24.56	& 20.44\\
  6&     Ca & 2.71	& 3.32	& 2.45\\
  7&     O  & 61.67	& 59.97	& 61.99\\
  8&     Sn & -	& 0.58	& -\\
\hline   

\end{tabular} 
\caption{Element compostion in the three glass samples in atomic weight percentage.}
\label{tab:EDS}  
\end{table}

\subsection{Construction of RPC Chambers}\label{sec:construction}
Using three different types of glasses mentioned in the previous section, we fabricated 
small prototypes of size 30 cm $\times$ 30 cm RPCs in our lab. Each  
glass plate electrode was chosen to be 3 mm and coated with a conductive layer of 
graphite on the outer surface. Buttons made up of polycarbonate material were glued 
on the electrode surface for the uniform gap separation of 2 mm between the two electrode plates. 
The gas gaps were sealed using polycarbonate side spacers from all sides. A gas inlet and 
a gas outlet nozzles made up of polycarbonate were glued on the diagonally opposite 
corners of the RPCs. The pickup panel consisted of honey comb panels with copper strips
 of width 2.80 cm placed above and below the electrodes.

After assembling the detector, we performed the leak test using manometer technique by measuring the pressure difference (water level difference in both arms)  with respect to time.

\section{Detector Characterization}\label{sec:char}
\begin{figure}
\centering
\includegraphics[height=7cm,width=15cm]{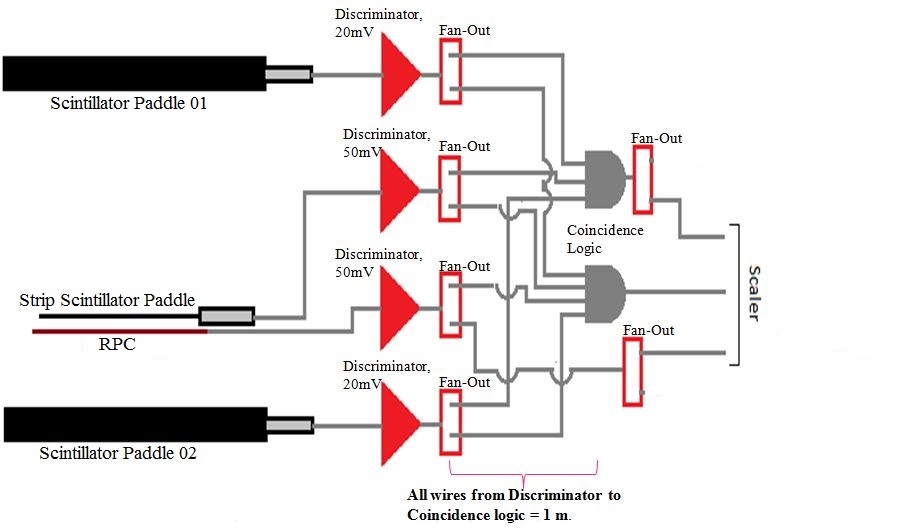}
\caption{ Setup for the muon telescope with DAQ}
\label{fig:DAQ}
\end{figure}

To characterize the performance of the RPCs made from the three types of glasses, we performed the efficiency, noise 
rate and leakage current measurements. A muon telescope consisting of three scintillator detectors being 
connected with the NIM/VME Data Acquisition system (DAQ) was used for conducting these measurements. The detector to be 
characterized is interleaved in between the scintillator detectors and readout by the 
DAQ system as shown in Fig.~\ref{fig:DAQ}. All the measurements reported in the next sections were performed at the normal pressure, relative humidity varying between 45\% to 50\% and temperature varying between $20^{\circ}C$ to $21^{\circ}C$ except for the temperature variation study reported in sec.~\ref{sec:Temp}.

The RPCs are characterized for efficiency, leakage current and noise rate under different environmental and running conditions. We 
characterize the RPCs for various gas compositions, temperatures and 
thresholds to obtain the optimum parameters to maximize the detector performance.

\subsection{Variation of Gas Mixture Composition}
The RPC detector uses different composition of gas mixtures to optimize the efficiency and longevity of the detectors. However, it has been found that choosing the best composition of gases is not an easy job and depends on many factors, like the material of the electrodes, the readout mode, the environment, etc. Many of the standard gas mixtures have unpleasant features such as flammability, formation of unwanted elements, etc. A large number of studies have already been performed in this direction~\cite{gas_Mengucci}-\cite{gas_Abe}. Therefore, it is very important to make an extensive study in order to choose the right composition of gases for the INO RPC detectors. 
The RPCs were tested in avalanche mode with four different gas mixture compositions as follows:

{\begin{enumerate}
\item First Composition : $R134a$ (95.0\%), $C_{4}H_{10}$ (5.0\%), $SF_{6}$ (0.0\%).
\item Second Composition : $R134a$ (90.0\%), $C_{4}H_{10}$ (10.0\%), $SF_{6}$ (0.0\%).
\item Third Composition : $R134a$ (95.0\%), $C_{4}H_{10}$ (4.5\%), $SF_{6}$ (0.5\%).
\item Fourth Composition : $R134a$ (90.0\%), $C_{4}H_{10}$ (9.0\%), $SF_{6}$ (1.0\%).
\end{enumerate}}

\begin{figure}[H]
\begin{minipage}{\linewidth}
      \centering
      \begin{minipage}{0.3\linewidth}
          \includegraphics[width=5cm,height=6cm]{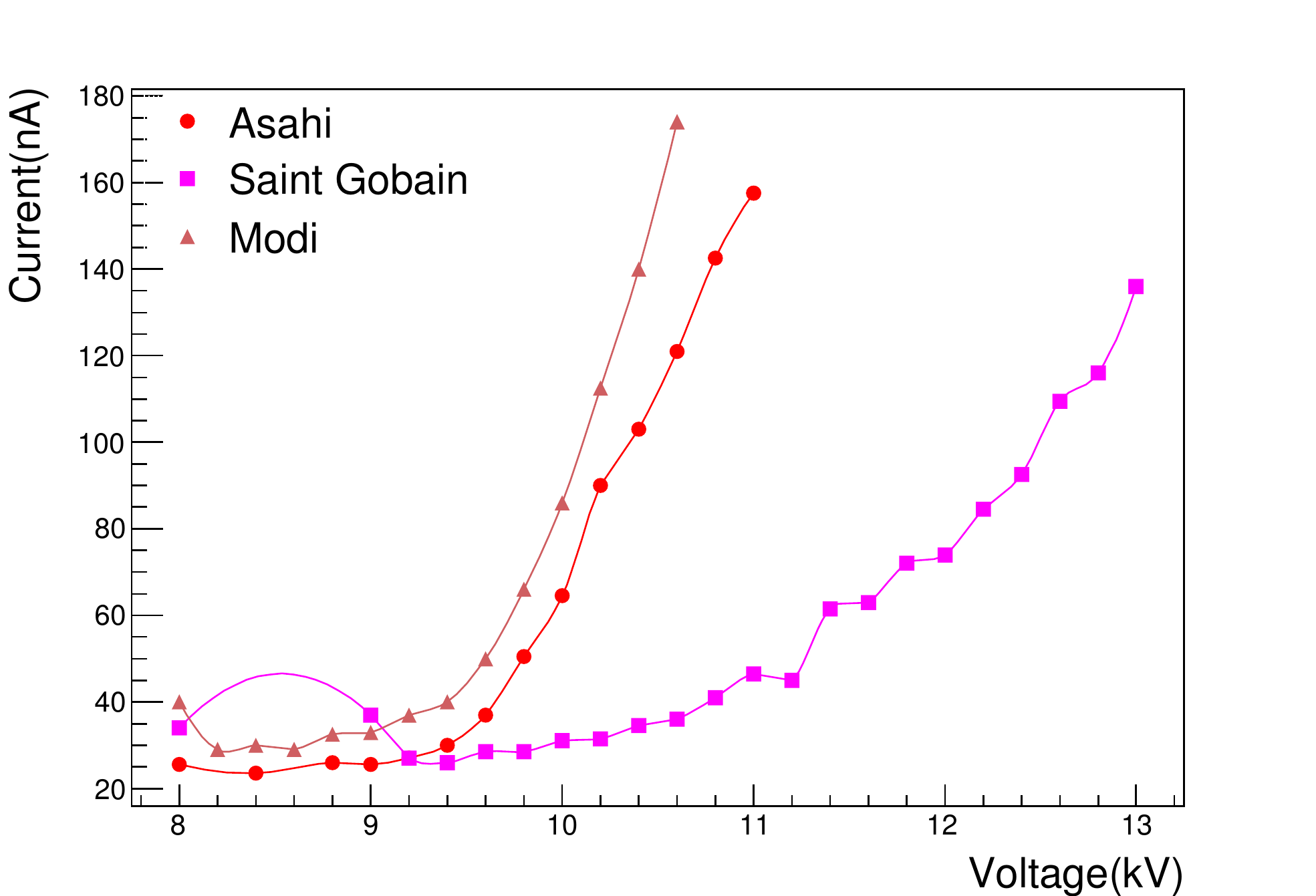}
               \end{minipage}
      \begin{minipage}{0.3\linewidth}
            \includegraphics[width=5cm,height=6cm]{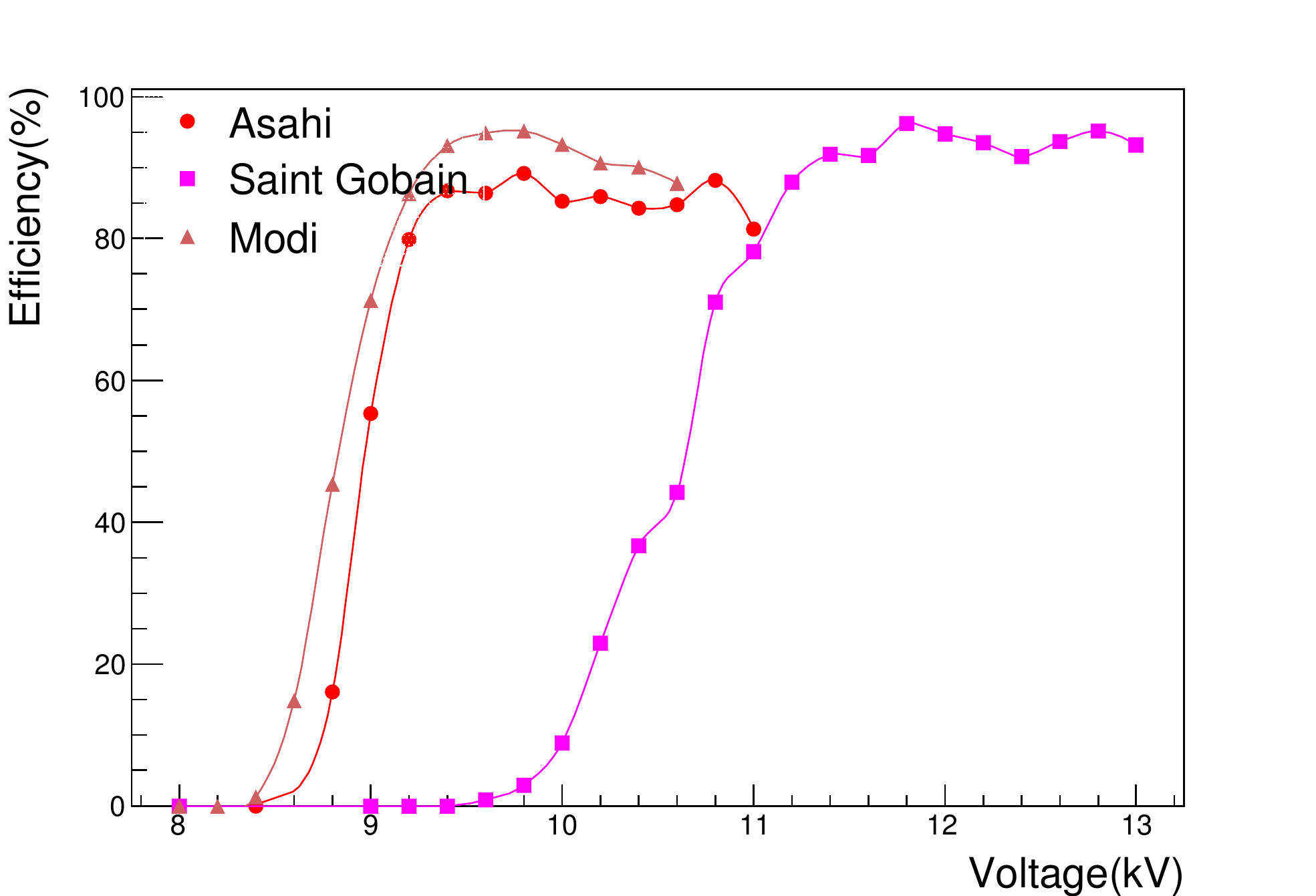}
             \end{minipage}
          \begin{minipage}{0.3\linewidth}
       \centering
            \includegraphics[width=5cm,height=6cm]{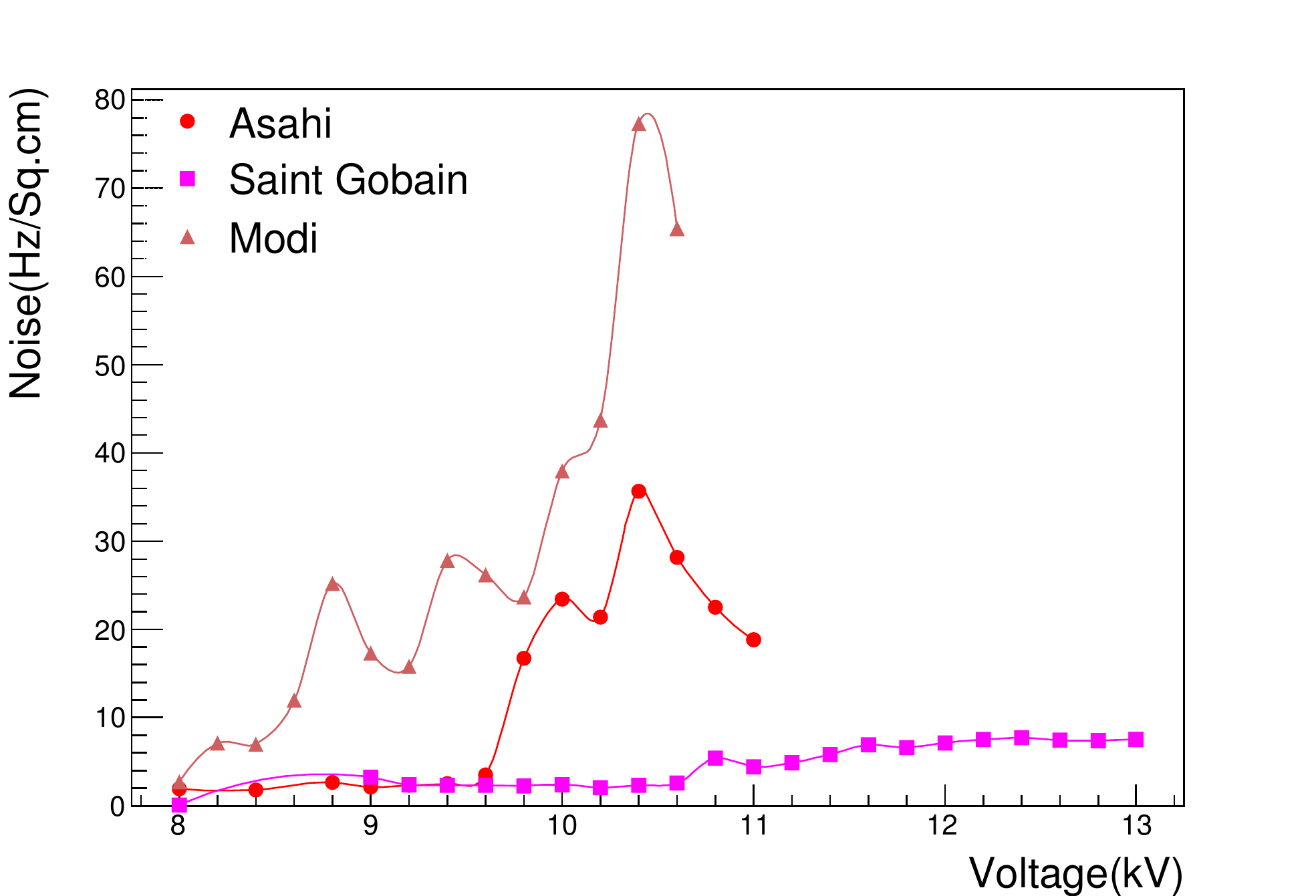}
        \end{minipage}
          \end{minipage}
          \caption{Leakage current, efficiency and Noise rate for the first composition.}
          \label{fig:GasMix1}
 \end{figure}

\begin{figure}[H]
\begin{minipage}{\linewidth}
      \centering
      \begin{minipage}{0.3\linewidth}
          \includegraphics[width=5cm,height=6cm]{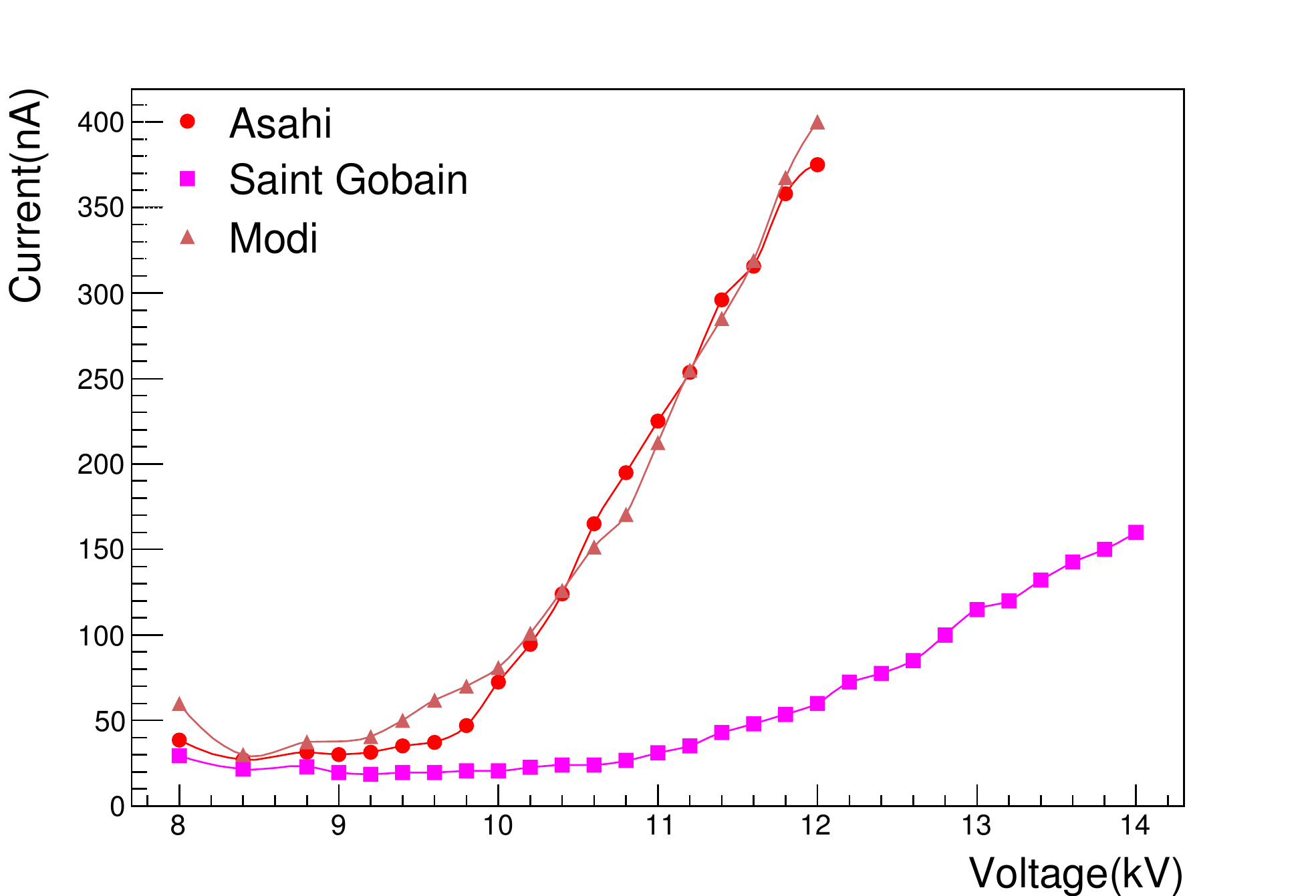}
               \end{minipage}
      \begin{minipage}{0.3\linewidth}
            \includegraphics[width=5cm,height=6cm]{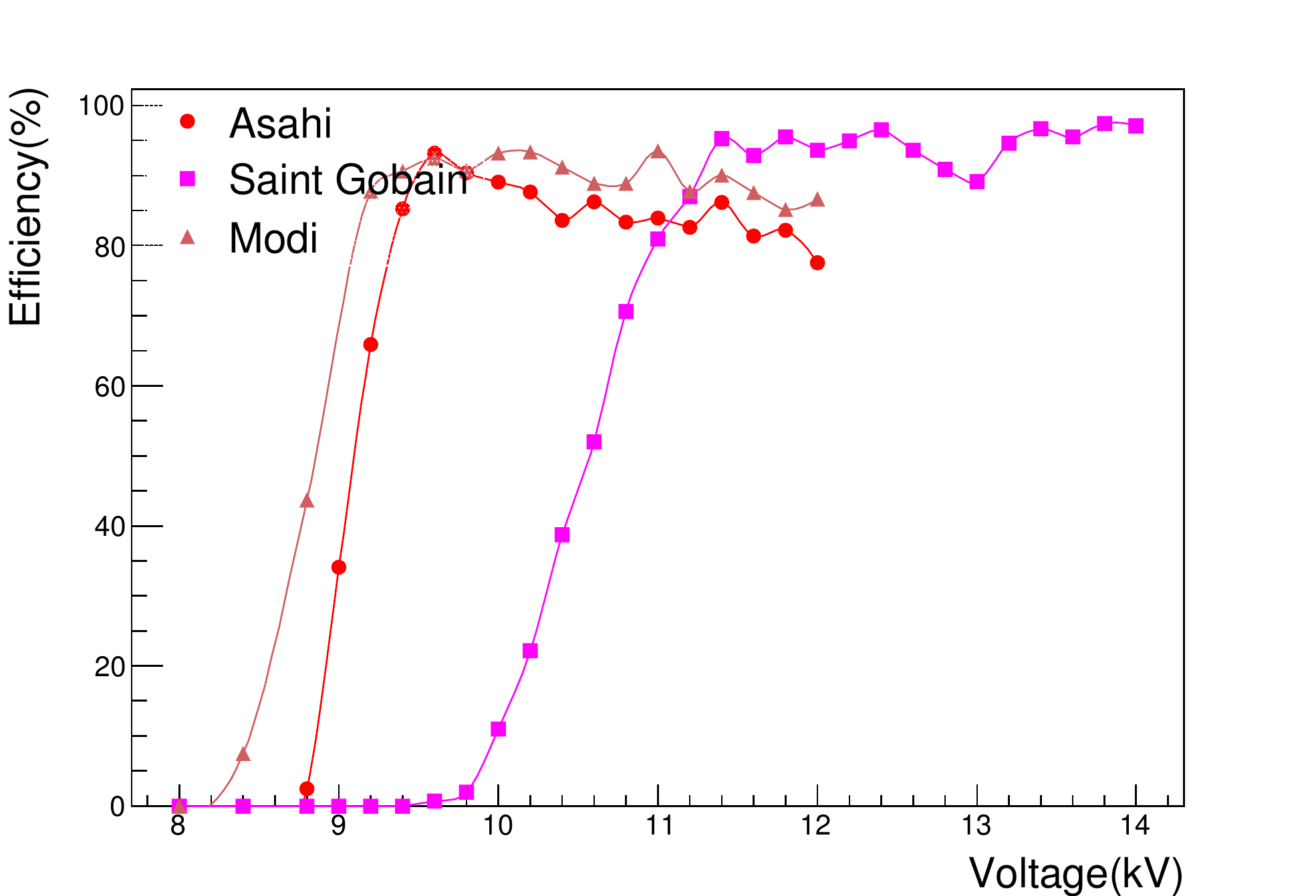}
             \end{minipage}
          \begin{minipage}{0.3\linewidth}
       \centering
            \includegraphics[width=5cm,height=6cm]{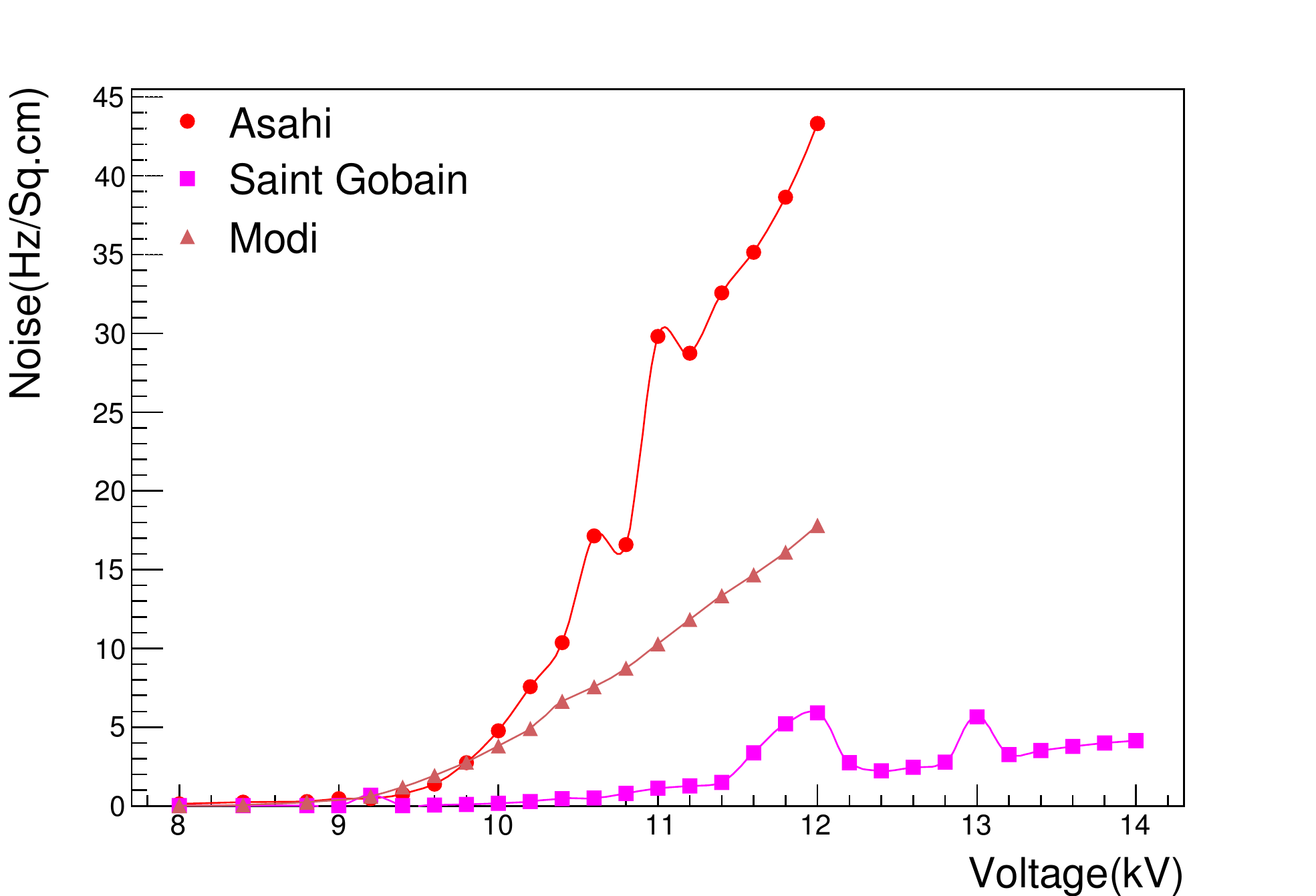}
        \end{minipage}
          \end{minipage}
          \caption{Leakage current, efficiency and Noise rate for the second composition.}
          \label{fig:GasMix2}
 \end{figure}

\begin{figure}[H]
\begin{minipage}{\linewidth}
      \centering
      \begin{minipage}{0.3\linewidth}
          \includegraphics[width=5cm,height=6cm]{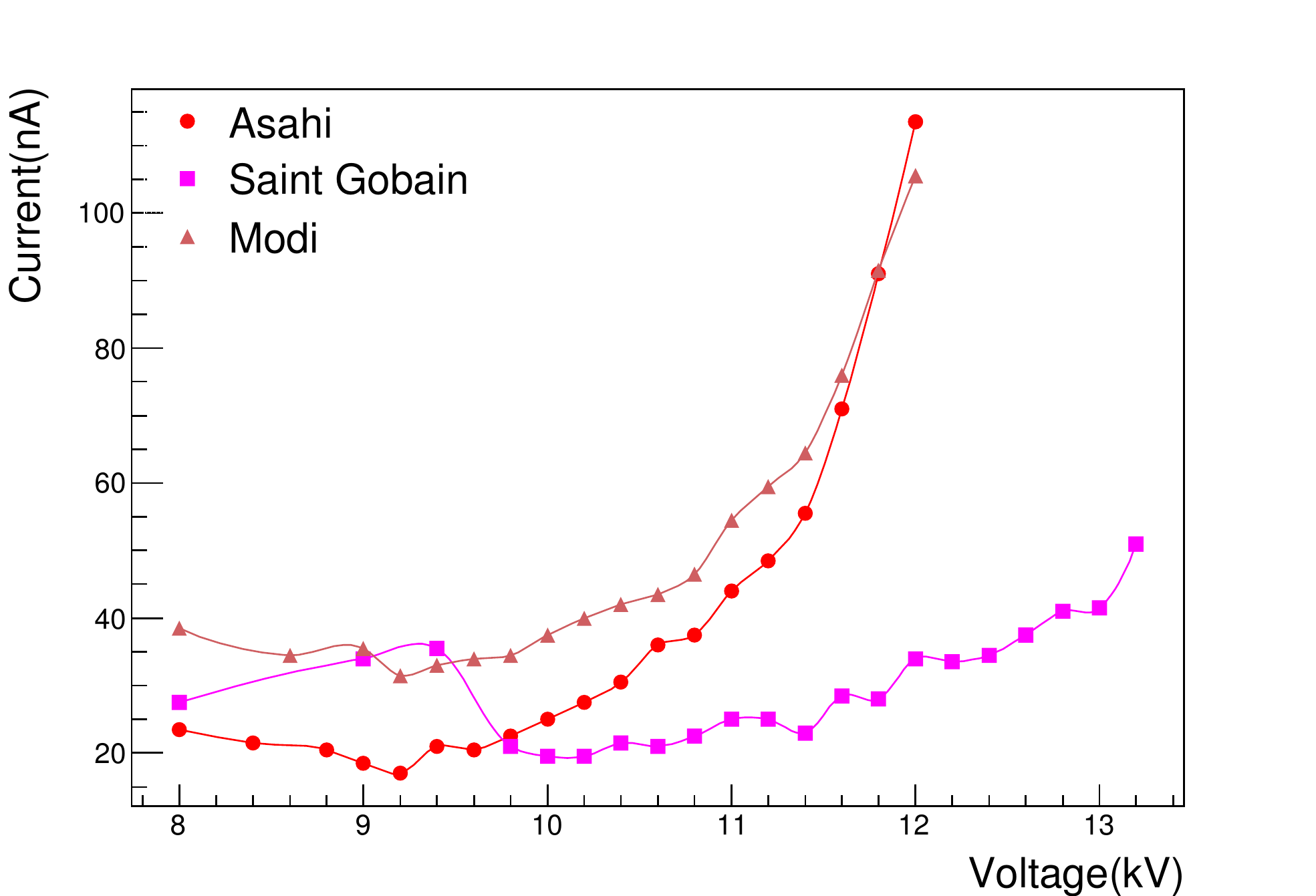}
               \end{minipage}
      \begin{minipage}{0.3\linewidth}
            \includegraphics[width=5cm,height=6cm]{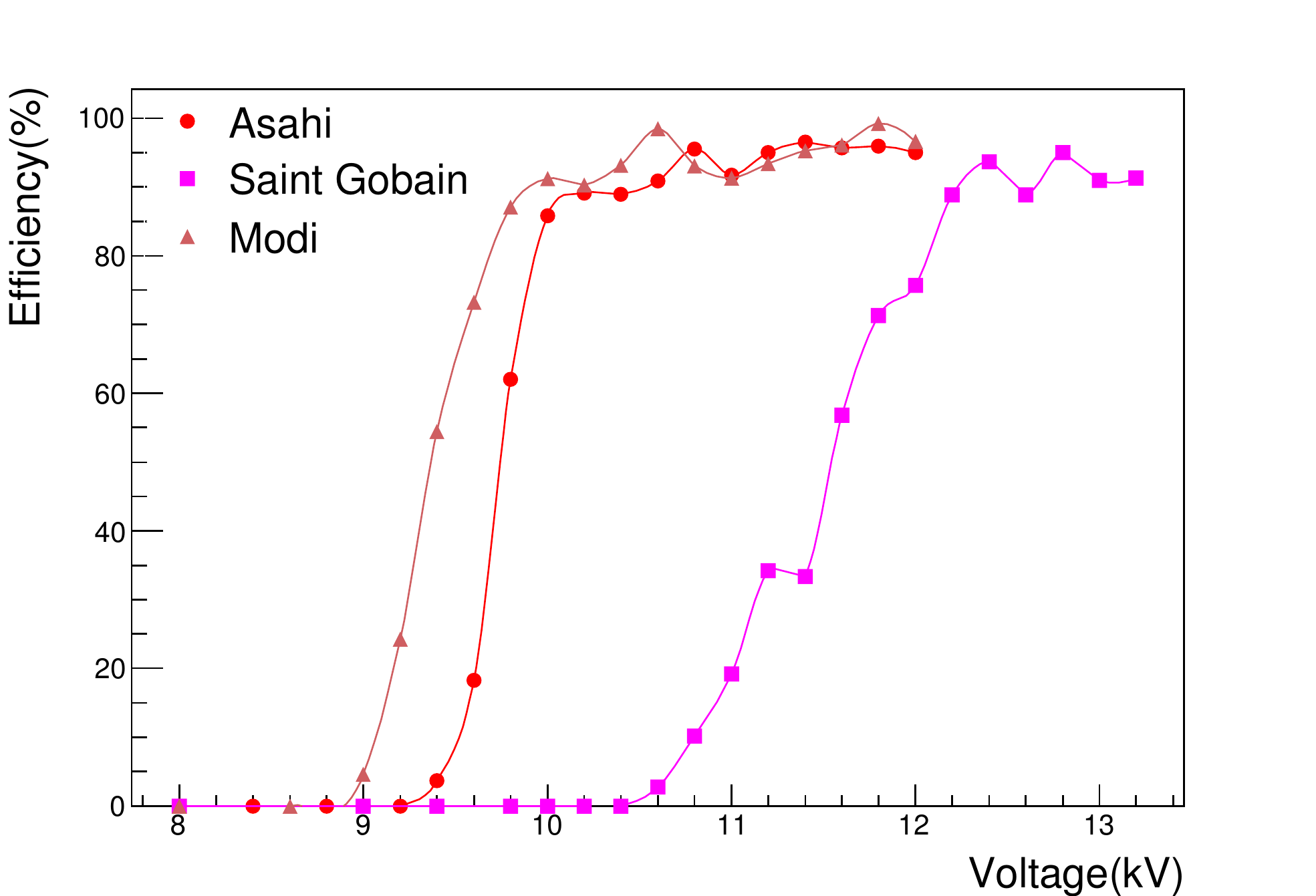}
             \end{minipage}
          \begin{minipage}{0.3\linewidth}
       \centering
            \includegraphics[width=5cm,height=6cm]{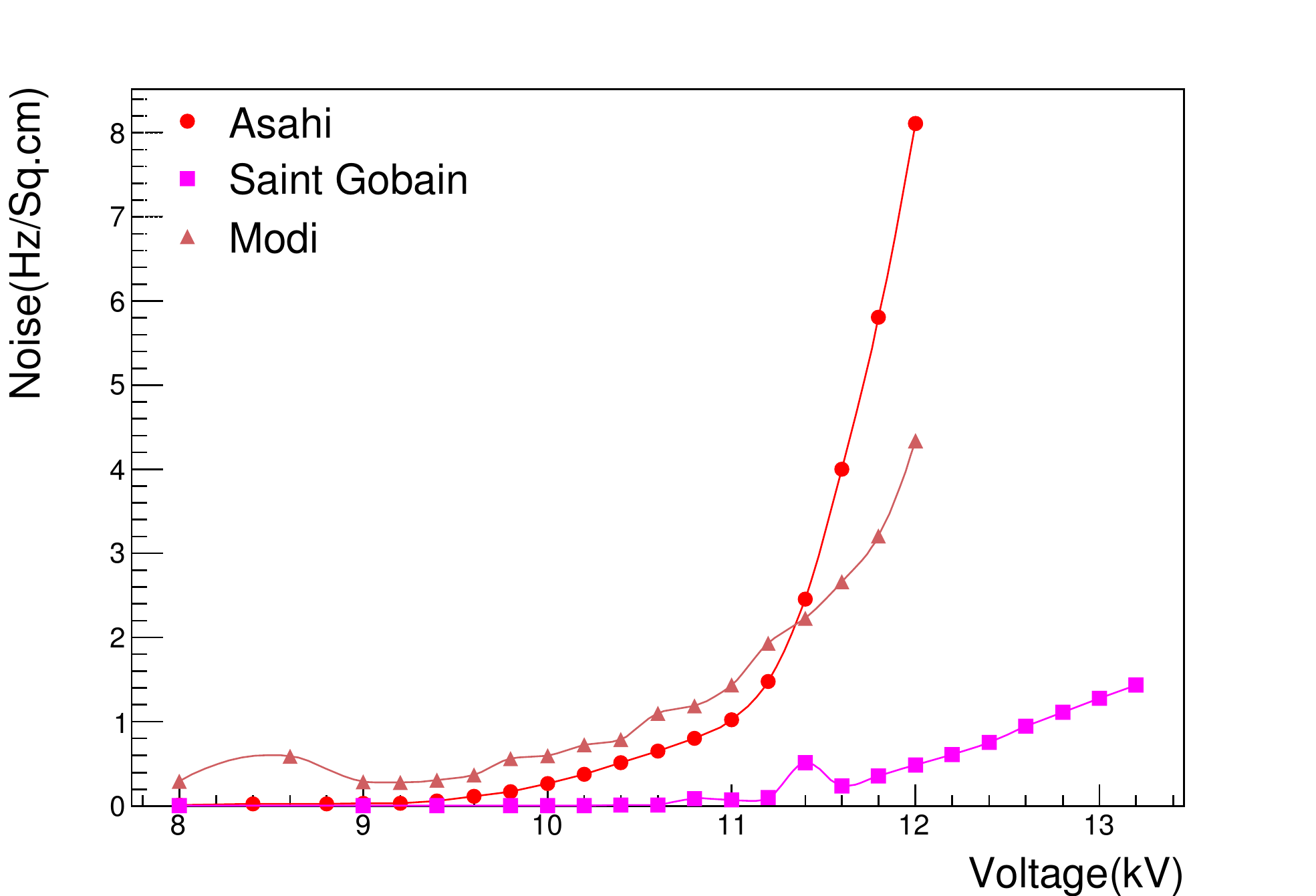}
        \end{minipage}
          \end{minipage}
          \caption{Leakage current, efficiency and Noise rate for the third composition.}
          \label{fig:GasMix3}
 \end{figure}

\begin{figure}[H]
\begin{minipage}{\linewidth}
      \centering
      \begin{minipage}{0.3\linewidth}
          \includegraphics[width=5cm,height=6cm]{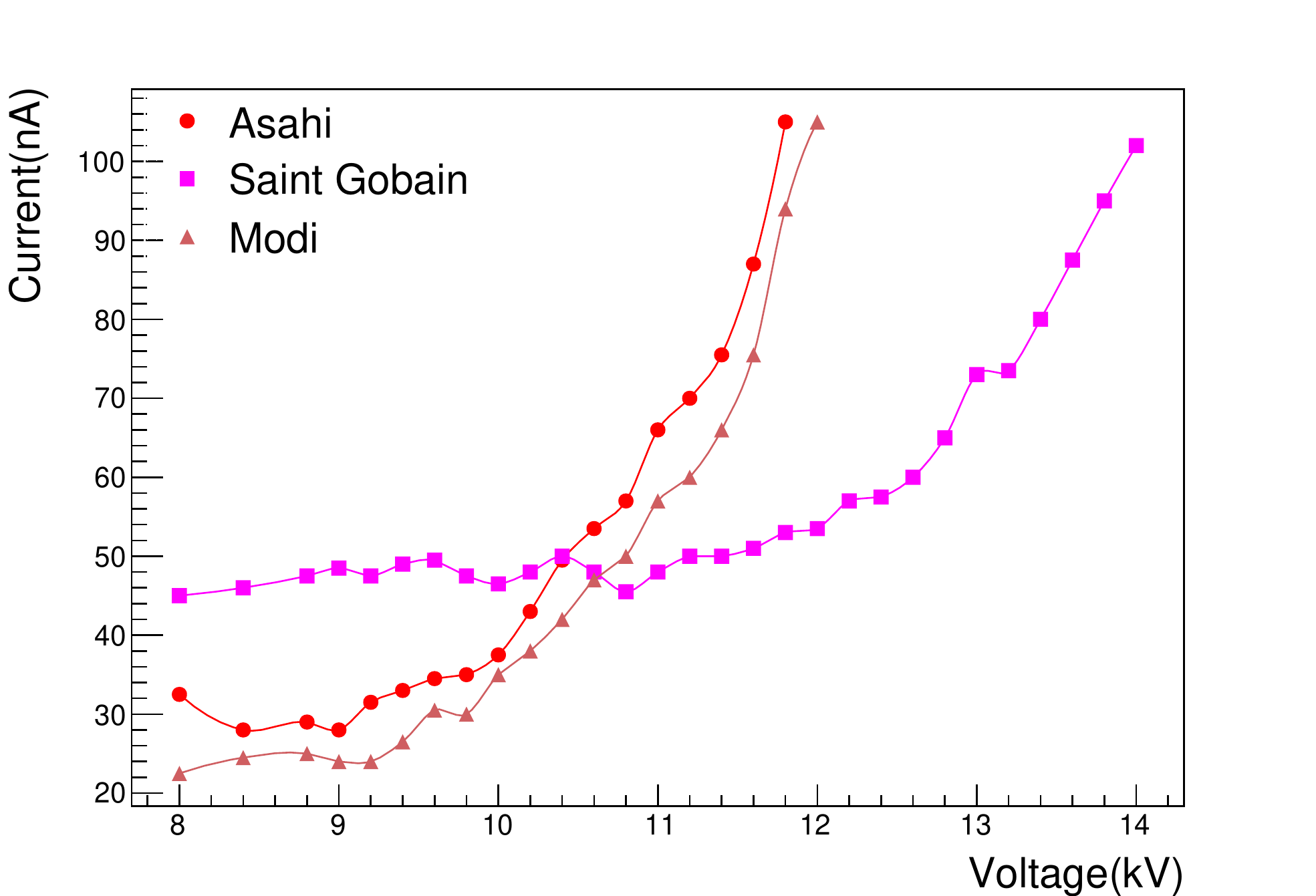}
               \end{minipage}
      \begin{minipage}{0.3\linewidth}
            \includegraphics[width=5cm,height=6cm]{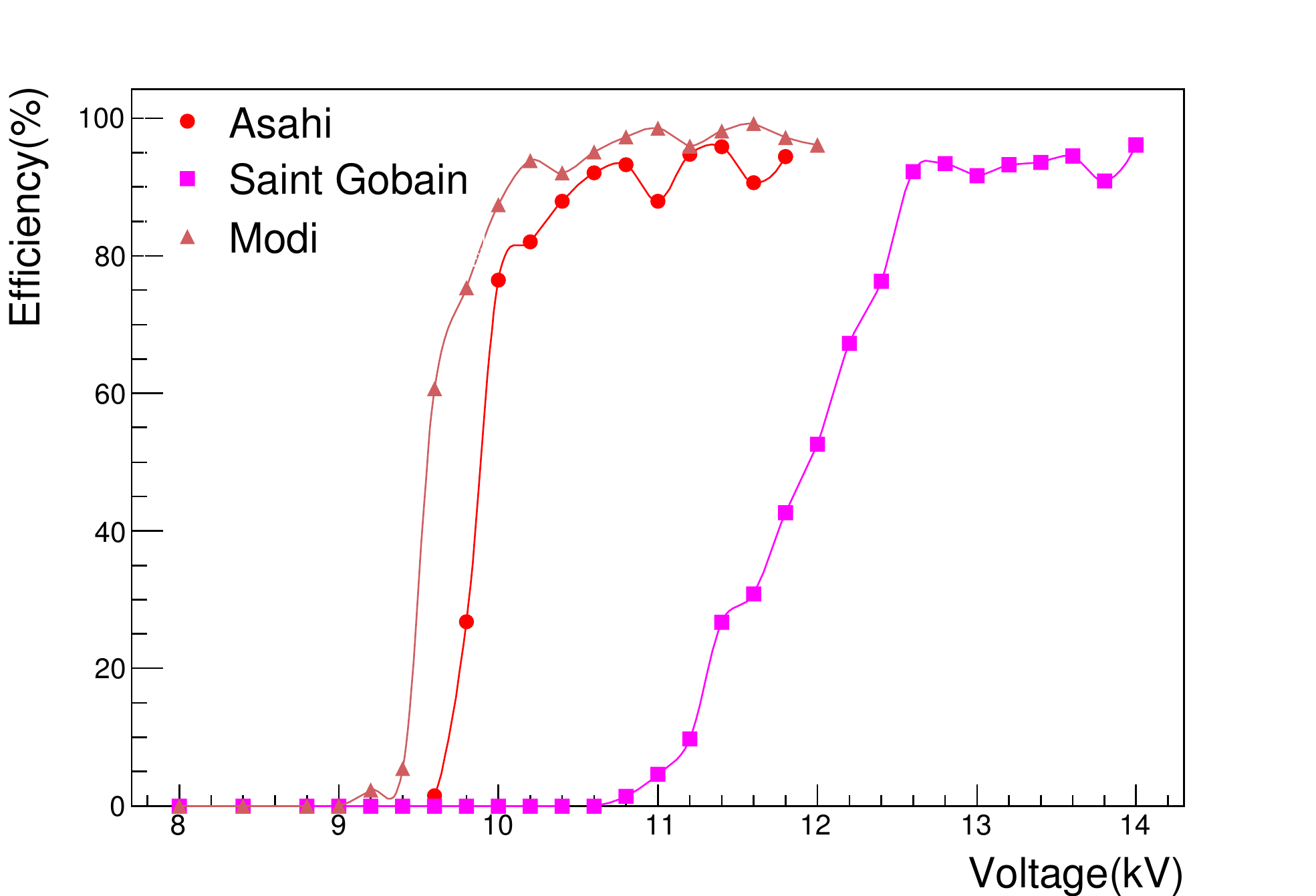}
             \end{minipage}
          \begin{minipage}{0.3\linewidth}
       \centering
            \includegraphics[width=5cm,height=6cm]{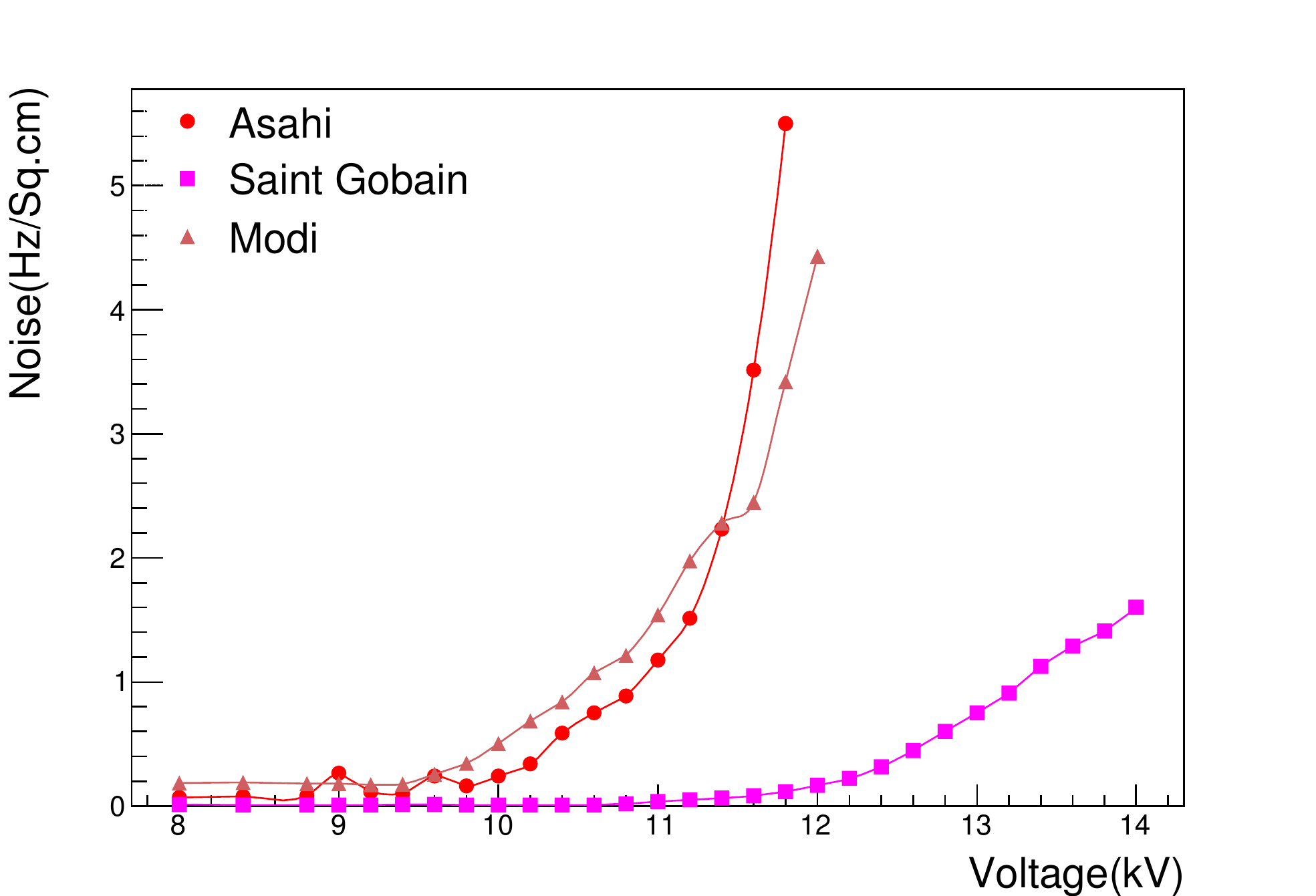}
        \end{minipage}
          \end{minipage}
          \caption{Leakage current, efficiency and Noise rate for the fourth composition.}
          \label{fig:GasMix4}
 \end{figure}

The flow rate was kept fixed at 10 SCCM (Standard Cubic Centimeter per Minute) for all the compositions. Since the Freon gas is the primary ionising gas so it would not make much sense to reduce the composition of this to a low level, therefore, we varied this between 90 and 95\%. The $C_{4}H_{10}$ is a flammable gas so it would not be advisable to increase its composition to large fraction due to safety reasons, therefore, we varied this between 4.5 to 10\%. The $SF_{6}$ gas is used as a quenching gas therefore,  its composition has been kept low and varied from 0 to 1\%. For all the above compositions, we measured the efficiency, noise rate and leakage current of the RPCs 
made from all the three types of the glasses. Fig.~\ref{fig:GasMix1} shows the leakage current, efficiency, and noise rate  
for the first gas composition. Fig.~\ref{fig:GasMix2}, Fig.~\ref{fig:GasMix3} and Fig.~\ref{fig:GasMix4} shows the same quantities for the second, third and fourth gas compositions respectively. From these figures it can be seen that the efficiencies 
for all these gas composition are similar for all the types of the glasses and is above 95\%. Noise rate as expected is maximum in the absence of $SF_{6}$ and decreases as we increase the $SF_{6}$ fraction. The noise rate is maximum for Modi and minimum for Asahi. Leakage current is much higher for Modi compared to Asahi and Saint Gobain which have reasonable values. These observations are consistent with the roughness measurement reported in  section~\ref{sec:surfaceProp}, where we found Modi glass having the most rough surface while Asahi having the least rough surface with Saint Gobain in between. 

\subsection{Variation of Thresholds} 

\begin{figure}[H]
\begin{minipage}{\linewidth}
      \centering
      \begin{minipage}{0.3\linewidth}
          \includegraphics[width=5cm,height=6cm]{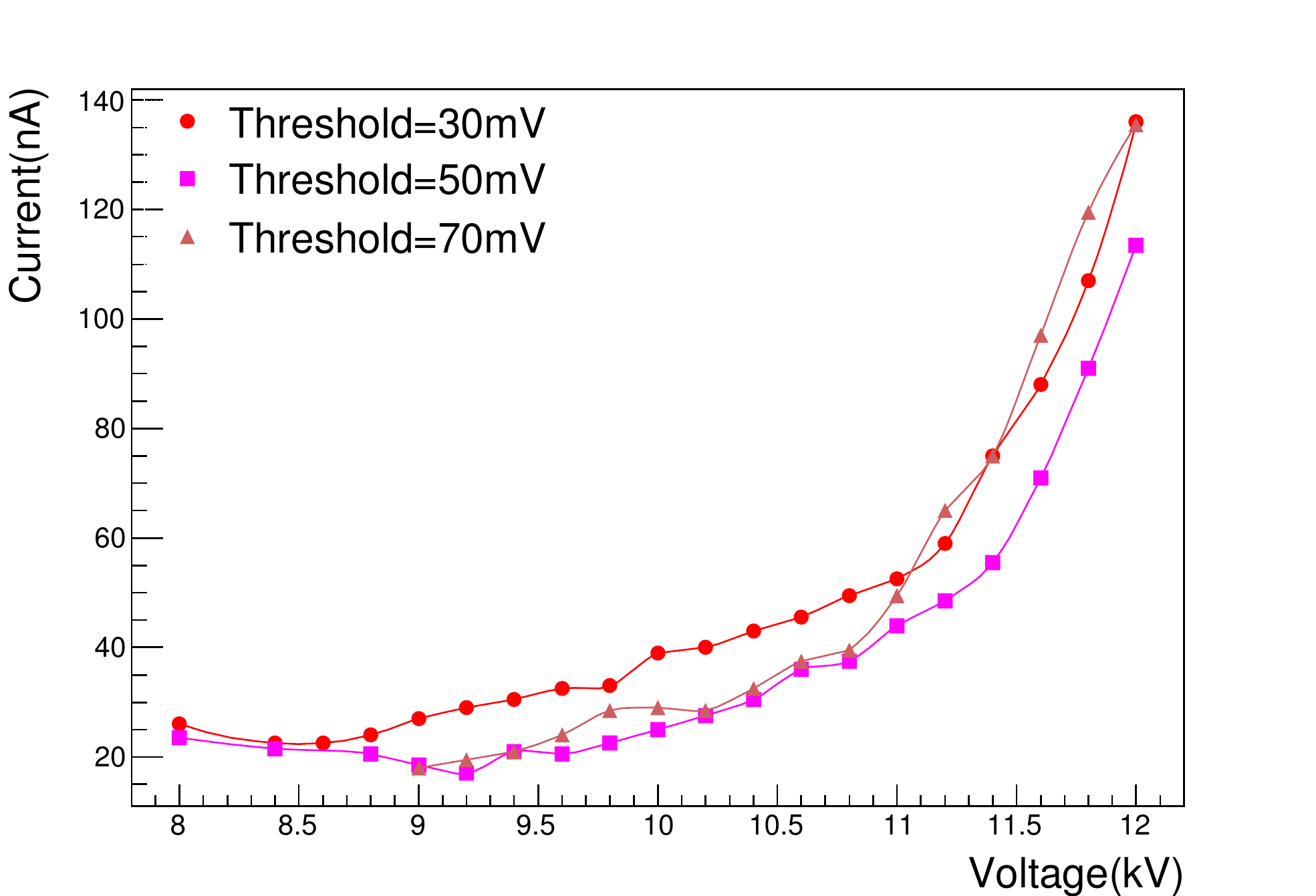}
               \end{minipage}
      \begin{minipage}{0.3\linewidth}
            \includegraphics[width=5cm,height=6cm]{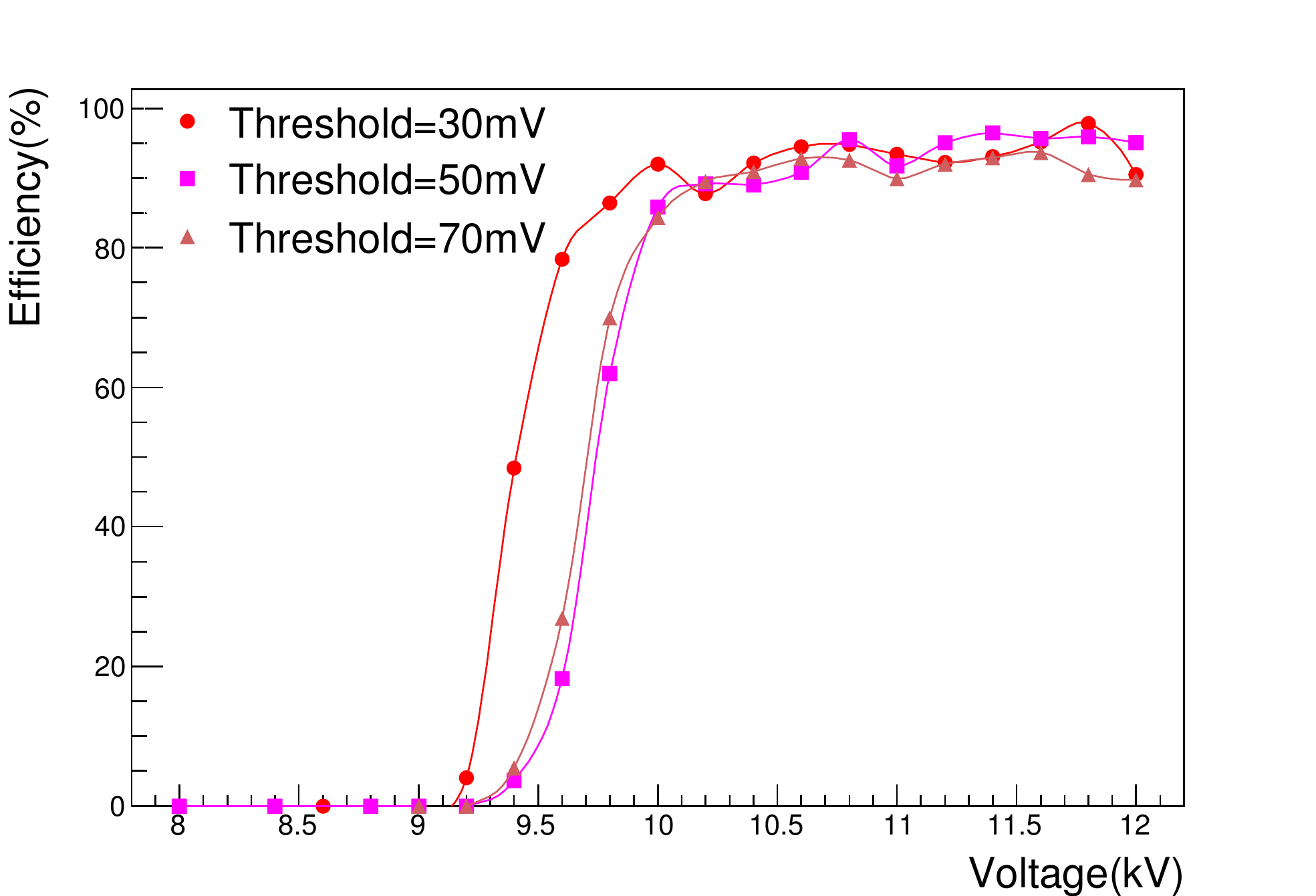}
             \end{minipage}
          \begin{minipage}{0.3\linewidth}
       \centering
            \includegraphics[width=5cm,height=6cm]{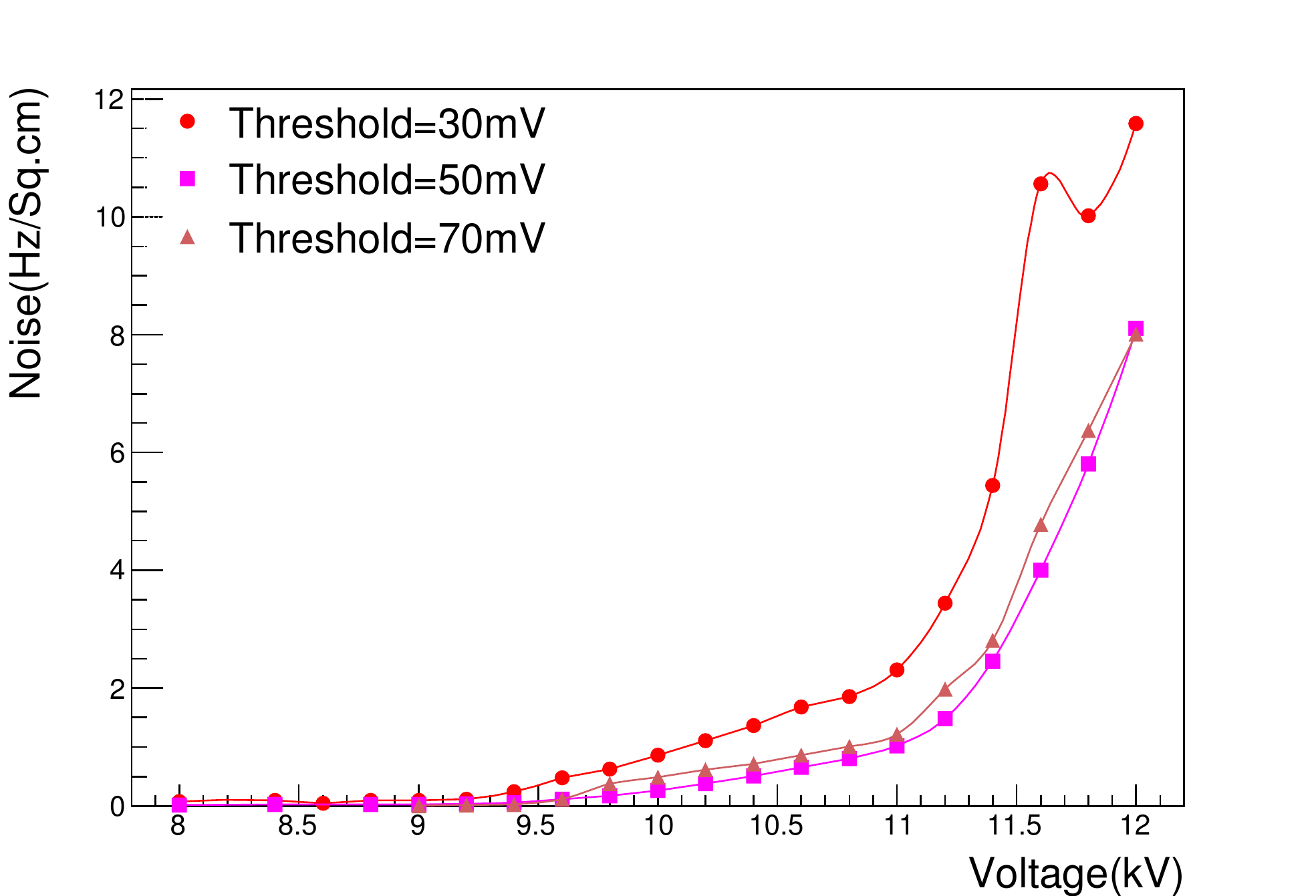}
        \end{minipage}
          \end{minipage}
          \caption{Leakage current, efficiency and Noise rate for Asahi Glass at three different threshold values and with third gas composition}
          \label{fig:Threshold}
 \end{figure}

The threshold value set in the discriminator circuit to reduce the spurious count rate may 
also affect the performance of the detector. In order to see the effect of this we 
varied the thresholds values for the RPC and scintillator paddles and measured the 
efficiencies and noise rates. Fig.~\ref{fig:Threshold} shows the effect of threshold variation on the efficiency and noise rates for Asahi glass RPC. We did not 
observe much dependence in efficiency between the threshold of 30 mV and 50 mV, whereas, the noise rate decreases as we increase 
the threshold value from 30 mV to 50 mV. However, further increasing the threshold 
value up to 70 mV did not decrease the count rate much, so we fixed the threshold value 
of 50 mV for further studies.

\subsection{Variation of environmental Temperature}
\label{sec:Temp}
\begin{figure}[H]
\begin{minipage}{\linewidth}
      \centering
      \begin{minipage}{0.3\linewidth}
          \includegraphics[width=5cm,height=6cm]{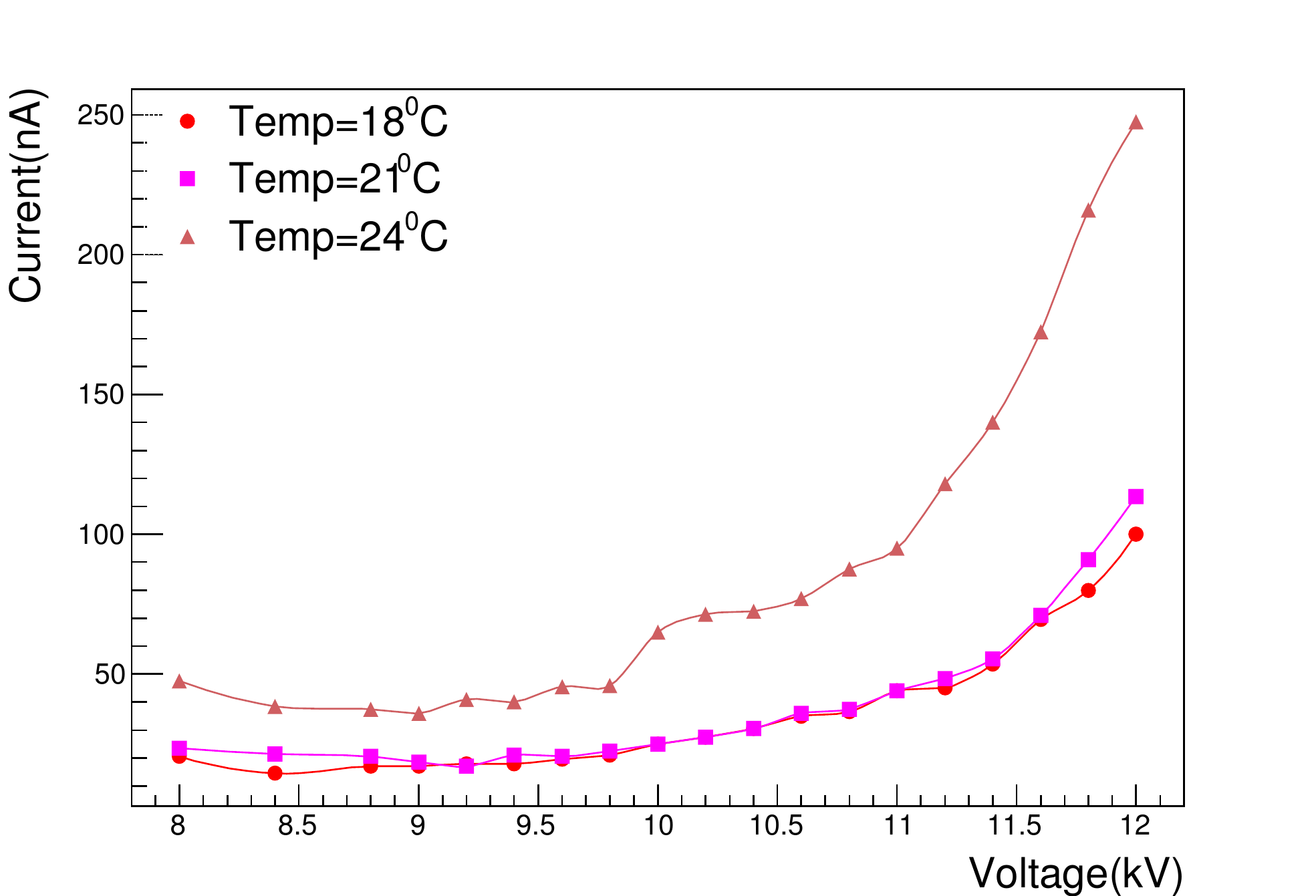}
               \end{minipage}
      \begin{minipage}{0.3\linewidth}
            \includegraphics[width=5cm,height=6cm]{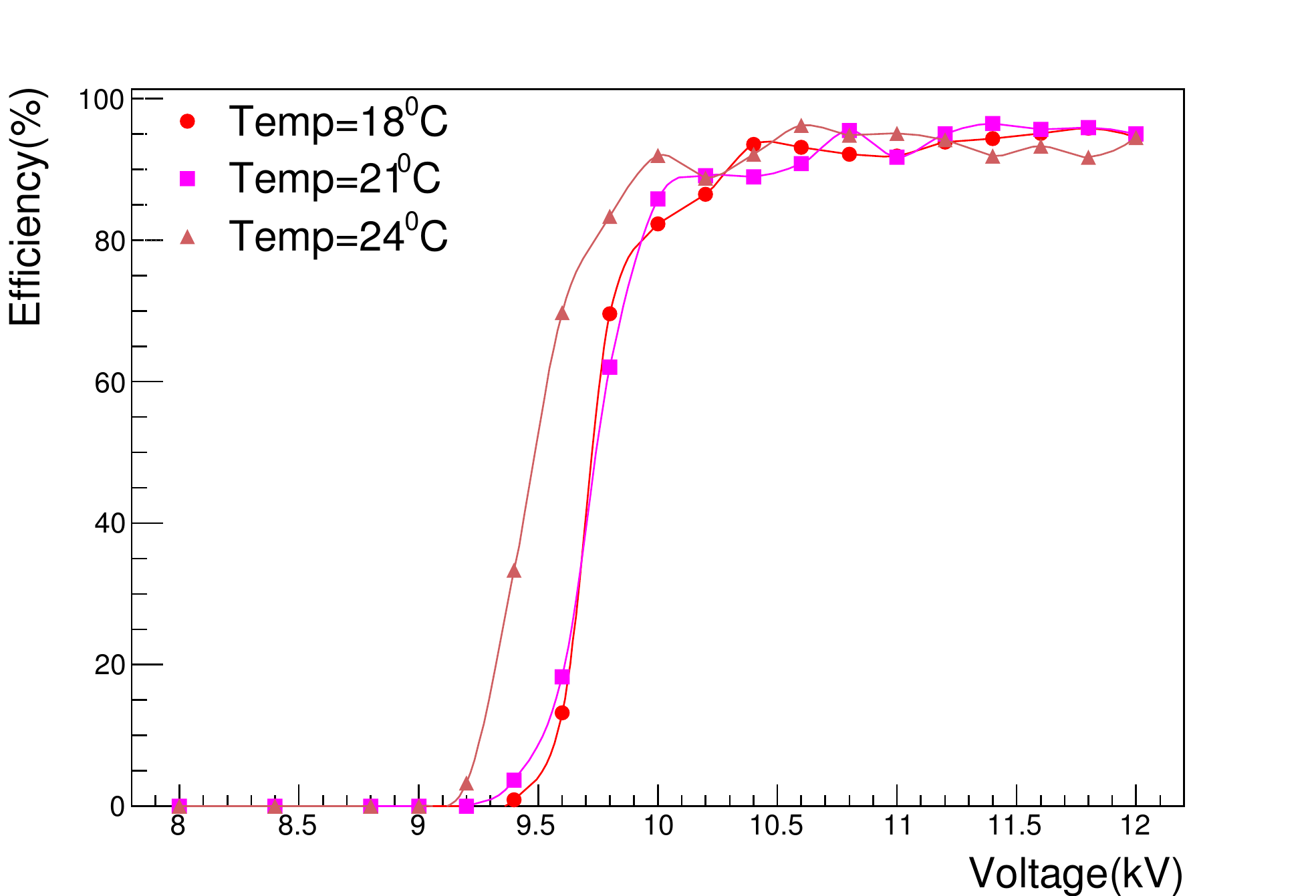}
             \end{minipage}
          \begin{minipage}{0.3\linewidth}
       \centering
            \includegraphics[width=5cm,height=6cm]{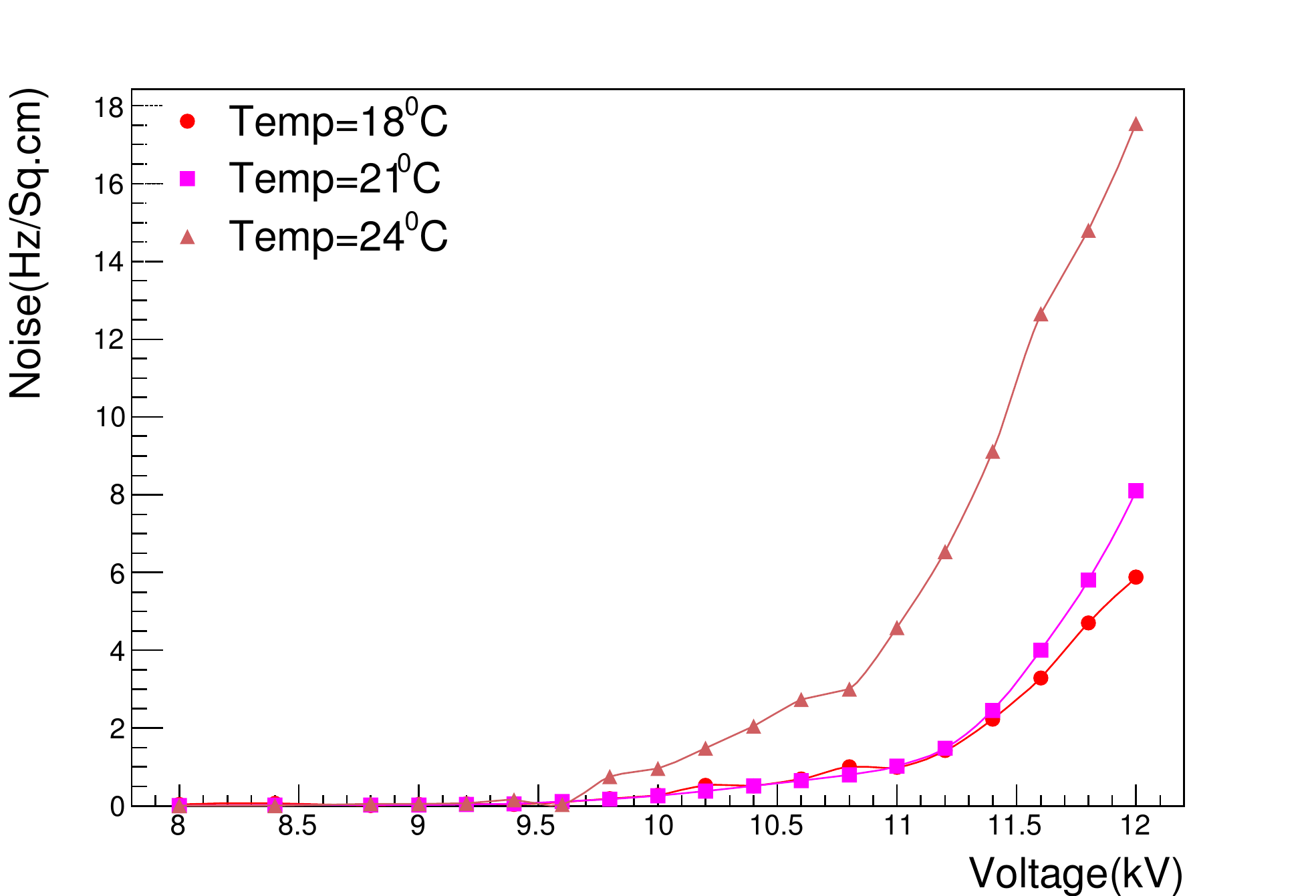}
        \end{minipage}
          \end{minipage}
          \caption{Leakage current, efficiency and noise rate for Asahi Glass at three different temperatures. These measurements are with third gas compositions. }
          \label{fig:Temp}
 \end{figure}

Environmental conditions like temperature and moisture also affects the performance 
of the RPC detectors, so it is very important to find out the suitable environmental 
conditions to operate the RPC in order to optimize the performance. We varied the 
temperature from $18^{\circ}$ C to $24^{\circ}$C and measured the efficiency, noise rate and 
leakage current for the RPCs made up from all the three types of glasses. Fig.~\ref{fig:Temp} shows the effect of temperature variation on the efficiency, current and noise rate for the RPC made up of Asahi glass. Other two glasses also show the similar characteristics.

\section{Results and Conclusions}\label{sec:res}
The INO-ICAL experiment will be using approximately 28,000 RPCs of size 2 m $\times$ 2 m. 
Before we embark up on building such a large amount of RPCs it is important to perform 
a thorough $R\& D$ on all aspects of the detector performance before freezing various 
parameters. We choose three types of glasses available in the local 
market, {\it{viz.}} Asahi, Saint Gobain and Modi. We performed various studies to assess 
the electrical properties of these three types of glasses. From our studies, we conclude 
that qualitatively Asahi is better than Saint Gobain and Modi in terms of smoothness. 
Saint Gobain is best in terms of bulk resistivity.

From performance studies, we find that all the three glasses gives almost similar efficiencies under all conditions. The threshold voltage for Saint Gobain is higher than others and the reason for this could be the higher bulk resistivity of the Saint Gobain glass. Saint Gobain shows lowest count rate for all the gas compositions. Leakage current is lowest for Saint Gobain for the first three gas compostions while Asahi shows the lowest current for the fourth gas composition. However, at lower bias voltages large fluctuation in the current was observed. The error on the current measurement is $2\% \pm$ 9~nA, so the fluctuation in current measuements at low voltages are consistent with each other within errors.
From our gas mixture composition studies, we find that the efficiency is not much dependent on the gas composition. 
The third composition of $R134a$ (95.0\%), $C_{4}H_{10}$ (4.5\%), $SF_{6}$ (0.5\%) is found to be giving the lowest noise rate and least current, therefore, we used this composition to perform further measurements.
 The study on variation of temperature shows that with increase in temperature from 18 to 21 $^{\circ}$C, there is not much effect on leakage current and noise rate. But, further increasing the temperature to $24^{\circ}$C, both the leakage current and noise rate increases. It is to be noted, however, that the relative humidity was consistently on the higher side (50\%) for the measurements done at the temperatures of 18$^{\circ}$C and 24$^{\circ}$C whereas it was on the lower side (around 45\%) for the measurements performed at the 21$^{\circ}$C. We believe this variation in atmospheric relative humidity to be the reason for transition in leakage current and noise rate at 24$^{\circ}$C when the humidity is also on the maximum side. We did not find any considerable effect of temperature variation on efficiency.

The discriminator threshold variations show that increasing the threshold from 30 mV 
to 50 mV decreased the spurious count rate considerably. However, further increasing the 
threshold from 50 mV to 70 mV did not have any effect on count rate. The efficiency 
does not depended much on the threshold variation from 30 mV to 50 mV. From this study we conclude that the 
threshold of 50 mV is a reasonable choice to operate the RPCs. 

In conclusion, we found that Saint Gobain and Asahi glasses are best suited for the INO-ICAL RPCs in terms of 
most of the parameters that we studied. We suggest to operate the RPC under the $R134a$ (95.0\%), $C_{4}H_{10}$ (4.5\%), $SF_{6}$ (0.5\%) gas composition and keep the temperature around $20^{\circ}$C in the INO cavern. It is to be noted, however, that all the studies reported in this paper were performed on small prototypes RPCs of size 30 cm $\times$ 30 cm. We are in the process of constructing actual size INO-ICAL RPCs to further continue our studies.

\section{Acknowledgments}

We would like to thank the Department of Science and Technology (DST), India and University of Delhi $R\&D$ grants for providing the financial support. Daljeet Kaur would like to thank Council of Scientific and Industrial Research (CSIR), India for providing the financial support. We would also like to thank the INO group at Tata Institute of Fundamental Research (TIFR) for providing 
some of the raw materials for detector construction. Our Sincere thanks to Prof. J. P. Singh of IIT Delhi for his help in getting AFM scans at IIT Delhi.

\end{document}